\documentclass[11pt]{article}

\usepackage[final]{acl}

\usepackage{times}
\usepackage{latexsym}
\usepackage{enumitem}
\usepackage{amsmath}
\usepackage{algorithm}
\usepackage{algpseudocode}
\usepackage{amsmath}
\usepackage{amsfonts}
\usepackage{pifont}
\usepackage[table]{xcolor}
\usepackage{tcolorbox}

\usepackage{booktabs}
\usepackage{longtable}
\usepackage{pdflscape}
\usepackage{array}
\usepackage{makecell}

\usepackage{adjustbox}
\usepackage{svg}

\usepackage{bm}

\usepackage[table]{xcolor}
\usepackage{pifont}
\newcommand{\cmark}{\textcolor{green!60!black}{\ding{51}}}
\newcommand{\xmark}{\textcolor{red}{\ding{55}}}
\newcommand{\partialmark}{\textcolor{black}{\(\sim\)}}
\newcommand{\absent}{\textcolor{red}{Absent}}
\newcommand{\notexplicit}{\textcolor{orange!85!black}{\(\sim\) Not explicit}}

\usepackage{natbib}
\usepackage{tikz}
\usepackage{hyperref}
\usepackage[percent]{overpic}

\providecommand{\xmark}{\ding{55}}
\providecommand{\xmark}{\ding{55}}
\providecommand{\cmark}{\ding{51}}

\usepackage[T1]{fontenc}

\usepackage[utf8]{inputenc}

\usepackage{microtype}

\usepackage{inconsolata}

\usepackage{graphicx}

\usepackage{tcolorbox}
\usepackage{booktabs}
\usepackage{multirow}

\usepackage{tabularx}
\usepackage{multirow}
\usepackage[table]{xcolor}
\usepackage{array}

\title{Puzzles to Paradigms: A Survey of Human--Agent Collaboration in Scientific Innovation}

\author{
  \textbf{Zhenni Liang\textsuperscript{1}} \quad
  \textbf{Haoxuan Zhang\textsuperscript{1}} \quad
  \textbf{Haihua Chen\textsuperscript{1,*}} \\
  \textbf{Ting Xiao\textsuperscript{1}} \quad
  \textbf{Junhua Ding\textsuperscript{1}} \quad
  \textbf{Yang Zhang\textsuperscript{1,*}} \\
  \textsuperscript{1}University of North Texas, Denton, TX, USA \\
  \texttt{\{zhenniliang, haoxuanzhang\}@my.unt.edu} \\
  \texttt{\{haihua.chen, ting.xiao, junhua.ding, yang.zhang\}@unt.edu} \\
  \textsuperscript{*}Corresponding authors
}

\newcommand{\coordcell}[1]{#1}

\begin{document}
\maketitle

\begin{abstract}
Agentic AI systems are increasingly embedded in scientific workflows, including literature synthesis, hypothesis generation, tool use, writing, and peer review. However, existing surveys and evaluations often emphasize task performance while underexamining how humans and agents coordinate under different epistemic conditions of scientific work. This survey introduces \textbf{CSCW$\times$Kuhn}, a coordination-aware framework that connects computer-supported cooperative work dimensions with distinct scientific contexts. Using this framework, we synthesize recent human--agent systems and benchmarks to examine how coordination demands and failure risks may vary across contexts. Our synthesis identifies five recurring failure modes: procedural drift, grounding substitution, evaluative flattening, proxy optimization, and epistemic black-boxing, and argues for evaluations that move beyond task accuracy and toward grounding, provenance, disagreement preservation, and long-horizon scientific reliability. Our corpus and coding artifact are available at \url{https://github.com/jenlia024/CSCWxKuhn-artifact}.
\end{abstract}

\section{Introduction}

Agentic AI systems are increasingly entering scientific research workflows. These recent systems can retrieve and synthesize literature, generate hypotheses, operate scientific tools, write and revise manuscripts, simulate peer review, coordinate multiple agents, and execute data-analysis or experimental procedures \citep{wang2023scientific, chen2025ai4research, wei2025agentic, Zheng2025Automation, wang2024autonomousagentsurvey, guo2024multiagentsurvey,zhang2026evolving}. Apart from the automation of scientific tasks, these emerging developments are also reshaping how scientific work is organized, monitored, and evaluated. Scientific innovation is rarely a single-step problem \citep{kuhn2012}---it requires researchers to frame problems, compare evidence, maintain provenance, interpret uncertainty, revise assumptions, and decide whether a result is worth pursuing. As agentic systems become increasingly involved in these workflows, task completion is no longer a sufficient measure of success; effective and reliable human--agent coordination under various epistemic conditions matters too.

This question of human--agent coordination is central, as scientific workflows vary widely in uncertainty, evidence standards, and evaluation criteria. Some tasks resemble puzzle-solving within established methods \citep{kuhn2012}, such as screening papers, executing analysis code, or following a protocol. Others involve evaluation of anomalies, where humans and agents must judge uncertain findings, contested claims, novelty, or reviewer disagreement. Yet others support paradigm-shift exploration, where the goal is to search for alternative concepts, distant analogies, or new research directions. These contexts impose different coordination demands. A system built for consistent, bounded execution may falter in contexts where evidence is contested; another built for exploration or discovery---generating plausible hypotheses---may be unreliable where provenance or validation paths remain opaque.

Computer-supported cooperative work (CSCW) offers a fruitful theoretical foundation for analyzing these situations. CSCW treats collaboration as a sociotechnical achievement that depends on articulation work, coordination mechanisms, shared understanding, and awareness of ongoing activity \citep{schmidt1992taking, schmidt1996coordination, clark1991grounding, dourish1992awareness, gutwin2002descriptive}. These concerns are central to scientific workflows, where researchers must maintain alignment in goals, evidence, methods, assumptions, intermediate artifacts, and responsibility over multiple stages \citep{olson2000distance, neang2021data}. 

When agents enter such workflows, they do not merely automate isolated tasks; they become part of distributed coordination arrangements. A CSCW perspective thus helps explain why task-level performance is insufficient: a system may produce a seemingly correct output while failing to preserve workflow state, evidential grounding, disagreement, provenance, or human accountability.

\textbf{Existing surveys} have made substantial progress in mapping AI systems for science, agentic architectures, and human--AI collaboration (Table~\ref{tab:survey_gap_comparison_full}). \textit{AI4Science surveys} organize systems along research pipelines, including literature review, hypothesis generation, experimentation, analysis, and writing \citep{chen2025ai4research, Zheng2025Automation, Tie2026AIScientists}. \textit{Agentic-system surveys} analyze planning, tool use, memory, retrieval, and multi-agent coordination \citep{wang2024autonomousagentsurvey, guo2024multiagentsurvey, tran2025multiagentcollaboration}. \textit{Human--agent collaboration surveys} examine role allocation, agency distribution, interaction patterns, and control with human involvement \citep{gomez2024notcollaborative, vats2025lfm, zhang2025agency, Zou2026LLMHAS}. Although several surveys address the scientific epistemic context (Table~\ref{tab:survey_gap_comparison_full}), none explicitly link it to coordination demand and failure-mode analysis within a unified framework. In parallel, current evaluation methods often emphasize bounded task success, output correctness, or benchmark performance \citep{Lou2025AAAR, Chen2025ScienceAgentBench, Majumder2025DiscoveryBench, Wijk2025REBench}, while giving less attention to workflow integrity, evidential grounding, disagreement preservation, provenance, and accountability \citep{xiong2025truthhypo, Jin2024AgentReviewEP, latona2025aireviewlottery}.

In this survey, we address the analytical coverage gap by introducing \textbf{CSCW$\times$Kuhn}, a framework for analyzing human--agent collaboration in scientific innovation. We use the term ``human--agent collaboration'' for systems in which agents contribute to scientific workflows (e.g., planning, judgment, execution, revision) and where human decisions remain relevant for final scientific output. Drawing on CSCW, we characterize coordination demand through articulation work, common ground, and awareness. Using Kuhn's account of scientific change \citep{kuhn2012}, we distinguish three epistemic modes: puzzle-solving in normal science, anomaly evaluation, and paradigm-shift exploration. Together, these perspectives ground our analysis.

\begin{figure*}[t]
    \centering
    \includegraphics[width=0.85\linewidth]{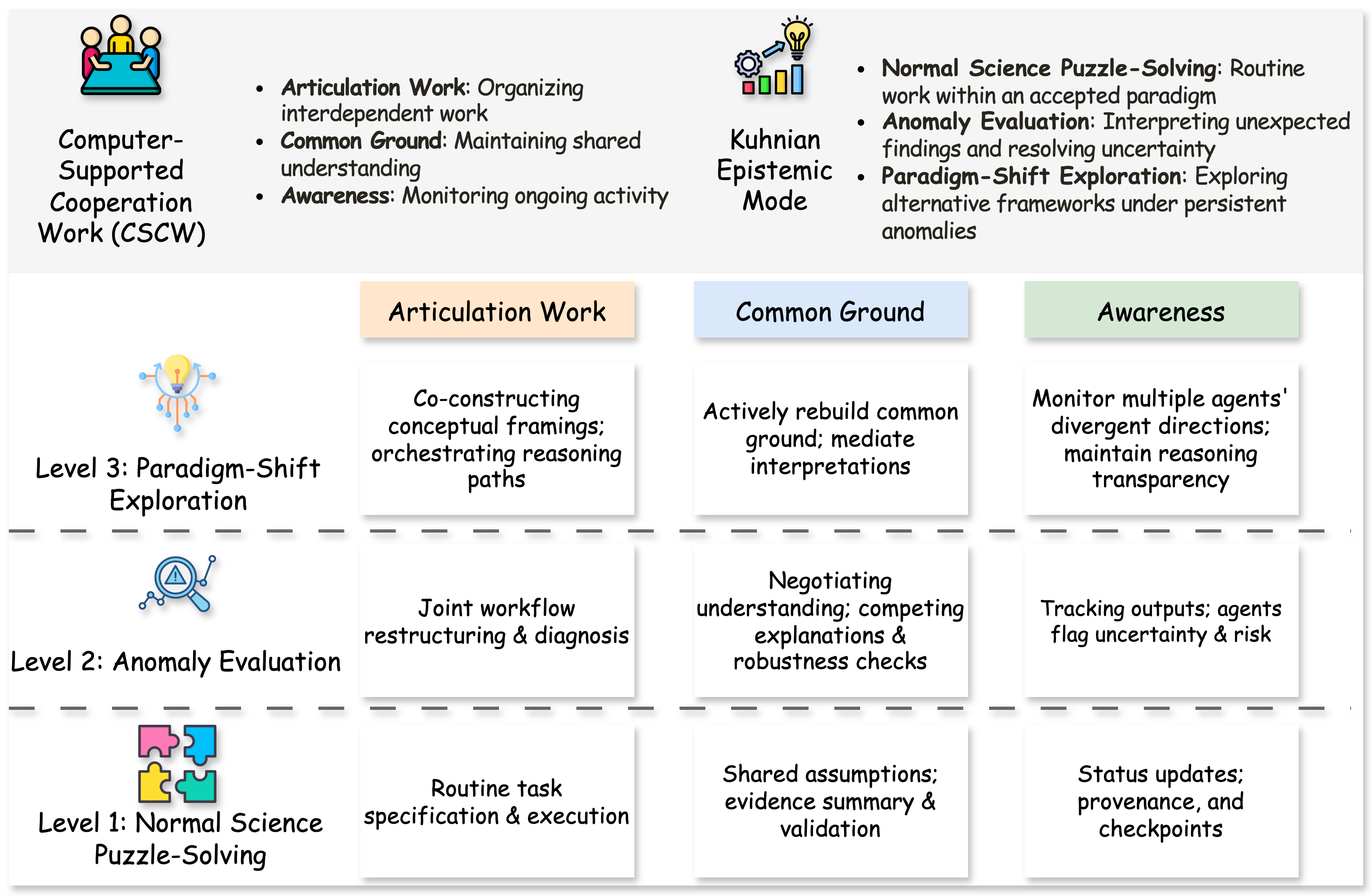} 
   \caption{CSCW$\times$Kuhn framework for human--agent scientific collaboration. It maps CSCW coordination dimensions to epistemic modes of scientific innovation, and proposes that higher epistemic uncertainty is associated with greater coordination, grounding, transparency, and oversight demands.}
    \label{fig:cscw_kuhn_matrix}
\end{figure*}

CSCW$\times$Kuhn proposes that coordination failure risks may emerge when the structure of human--agent collaboration is misaligned with the epistemic demands of the scientific task. Specifically, our synthesis suggests that \textit{procedural drift} is salient in bounded execution workflows, \textit{grounding substitution} is critical in hypothesis generation and literature-based reasoning, and \textit{evaluative flattening} becomes consequential when agents participate in peer review or novelty assessment. This framing helps explain why task-level success can still mask failures in scientific reliability. Figure~\ref{fig:cscw_kuhn_matrix} summarizes our framework. Our survey makes the following three contributions:
\begin{itemize}[leftmargin=*]
    \item \textbf{A coordination-aware taxonomy.} We introduce \textbf{CSCW$\times$Kuhn}, a coordination-aware analytical framework and taxonomy that connects CSCW coordination dimensions with epistemic modes of scientific work.

    \item \textbf{A synthesis of recent systems.} We organize representative human--agent scientific systems across puzzle-solving in normal science, anomaly evaluation, and paradigm-shift exploration to illustrate how collaboration patterns and system risks vary by epistemic context.

    \item \textbf{An evaluation agenda.} We identify five recurring coordination failure modes: procedural drift, grounding substitution, evaluative flattening, proxy optimization, and epistemic black-boxing, and translate them into evaluation targets beyond task accuracy.
\end{itemize}

We use CSCW$\times$Kuhn to synthesize recent systems and benchmarks over three scientific contexts. Section~\ref{sec2} defines the framework and formalizes coordination demand. Section~\ref{sec3} applies the framework to representative human--agent systems across puzzle-solving, anomaly evaluation, and paradigm-shift exploration. Section~\ref{sec4} analyzes current benchmarks and evaluation methods, showing where task-centric evaluation under-measures coordination quality. Section~\ref{sec5} discusses implications for system design, evaluation, and governance, and Section~\ref{sec6} concludes. This survey is not an exhaustive systematic review; it is intended as a focused, representative conceptual survey. Appendix~\ref{corpuscoding} details the search, screening, and coding procedures.

\section{A Coordination Framework for Human--Agent Scientific Collaboration}
\label{sec2}

Human--agent collaboration in scientific innovation is more than assigning tasks to agents or evaluating final outputs. Scientific work unfolds through multi-step workflows (e.g., problem framing, literature review, hypothesis generation, experimentation, analysis, review). In these workflows, humans and agents must coordinate goals, assumptions, evidence, intermediate states, and responsibilities. The required coordination varies across scientific contexts. Some tasks involve stable goals and accepted criteria; others involve uncertain claims and contested evidence; yet others require exploring entirely new conceptual directions.

We present CSCW$\times$Kuhn, a framework for analyzing these differences. Concepts from CSCW ground our framework in prior work on cooperative work and coordination. In CSCW, collaboration requires the organization of interdependent activities, the repair of workflow breakdowns, the maintenance of common understandings, and the visibility of ongoing work \citep{schmidt1992taking, schmidt1996coordination, clark1991grounding, dourish1992awareness, gutwin2002descriptive}. This perspective is useful because agent capabilities are deeply embedded in interdependent scientific workflows. Agents may retrieve evidence, decompose tasks, execute tools, critique claims, or revise outputs, while humans oversee, interpret scientific meaning, and decide if the resulting work is credible. Moreover, Kuhn's account of scientific change helps distinguish routine puzzle-solving, anomaly evaluation, and paradigm-shift exploration \citep{kuhn2012}. We combine CSCW and Kuhn to analyze how coordination demands and evaluation targets change in context, across human--agent scientific workflows.

\subsection{Coordination Dimensions}

We characterize coordination demand with three dimensions informed by CSCW:

\begin{itemize}[leftmargin=*]
    \item \textbf{Articulation work ($\alpha_t$)}: The work required to organize, assign, sequence, integrate, and repair interdependent tasks. This follows CSCW accounts of articulation work as the often-invisible organizing work that makes cooperative activity possible \citep{strauss1985work, schmidt1992taking}.
    
    \item \textbf{Common ground ($\gamma_t$)}: The shared understanding of goals, assumptions, constraints, evidence, methods, and task state. This dimension draws on grounding theory, which emphasizes that collaboration requires participants to establish and maintain sufficient mutual understanding for the activity at hand \citep{clark1991grounding}.
    
    \item \textbf{Awareness ($\omega_t$)}: The visibility of actions, intermediate outputs, decisions, errors, and consequences. This dimension follows CSCW work on awareness in shared workspaces, where participants need to understand what others are doing, what has changed, and how their own actions should adapt \citep{dourish1992awareness, gutwin2002descriptive}.
\end{itemize}

At workflow step $t$, we define coordination demand as:
\begin{equation}
    \phi_t = (\alpha_t, \gamma_t, \omega_t),
\end{equation}
where $\alpha_t$, $\gamma_t$, and $\omega_t$ correspond to articulation work, common ground, and awareness. This tuple is a qualitative, ordinal coding representation of what a human--agent system must support. Appendix~\ref{app:cscw-foundations} provides further discussion of this framing.

\subsection{Scientific Contexts}

Coordination demands depend on the scientific context. We use Kuhn's theory \citep{kuhn2012} as an analytical heuristic and distinguish three contexts:

\begin{itemize}[leftmargin=*]
    \item \textbf{Puzzle-solving in normal science}: Work conducted within accepted goals, methods, and evaluation criteria.
    
    \item \textbf{Anomaly evaluation}: Work that assesses unexpected findings, uncertain claims, or contested contributions.

    \item \textbf{Paradigm-shift exploration}: Exploratory work that searches for alternative concepts, cross-domain connections, or new research directions.
\end{itemize}

Let $K$ denote the set of scientific contexts:
\begin{equation}
K =
\left\{
\begin{array}{l}
\text{Puzzle-solving in normal science,} \\
\text{Anomaly evaluation,} \\
\text{Paradigm-shift exploration}
\end{array}
\right\},
\end{equation}
where $k_t \in K$ denotes the dominant context at step $t$. These categories are not strict boundaries; a scientific workflow may move from routine execution to anomaly evaluation or exploratory reasoning. Appendix~\ref{kuhnlen} provides a more in-depth discussion of this framing.

\subsection{Framework Formulation}

We model human--agent scientific collaboration as an iterative workflow that transforms an initial research intent into evidence-backed outputs. In systems with substantial agent--agent coordination, we treat these as an architectural feature of the broader sociotechnical workflow when humans retain control over objectives, evaluation, validation, or outputs, even if execution is autonomous.

Let $S = \{s_1, s_2, \ldots, s_n\}$ denote the set of workflow stages (e.g., problem framing, experimentation, review). The coordination demand at step $t$ depends on both the scientific context and the workflow stage:
\begin{equation}
    \phi_t = f(k_t, s_t),
\end{equation}
where $k_t \in K$ and $s_t \in S$. 

This formulation captures the main idea of CSCW$\times$Kuhn---different scientific contexts require different forms of human--agent coordination. In normal-science puzzle-solving, coordination mainly concerns reliable execution and tracking of workflow. In anomaly evaluation, it shifts toward evidential comparison, disagreement preservation, and evaluative alignment. In paradigm-shift exploration, coordination becomes more demanding because goals, assumptions, and validation criteria may be unstable.

Under this framework, workflow stage and epistemic mode are separate, distinct dimensions. The stage indicates where in the research workflow an activity occurs; epistemic mode characterizes the scientific conditions under which it occurs. We illustrate this independence with the following two examples.

Within puzzle-solving, literature review and protocol execution occupy different workflow stages but share a common epistemic mode, with stable goals and evaluation criteria. Their differing coordination demands thus stem from workflow stage, not from a shift in epistemic conditions. Conversely, literature-related work can occur under different epistemic modes at the same workflow stage. For instance, under puzzle-solving, routine evidence retrieval demands mainly reliable search and coverage, whereas under anomaly evaluation, conflicting literature assessment demands mainly evidential comparison and disagreement preservation.

Additionally, this framework clarifies why task-level evaluation is insufficient: a system may complete a benchmark task while failing to preserve workflow integrity or human accountability. Thus, we use CSCW$\times$Kuhn as a coding framework that links scientific context, coordination demand, failure mode, and evaluation target.

We operationalize CSCW$\times$Kuhn as a qualitative coding scheme. For each system or workflow segment, we code for the dominant epistemic context, the level of coordination support provided by the system along articulation work, common ground, and awareness, and the corresponding failure risks and evaluation targets. Appendix~\ref{app:coding-rubric} provides the full rubric and decision rules.

\section{Collaboration Patterns Across Scientific Contexts}
\label{sec3}

\paragraph{Corpus and Coding.}
To apply CSCW$\times$Kuhn, recent human--agent systems and benchmarks were coded along four dimensions: epistemic mode, system cluster, coordination demand, and potential failure risk. The corpus was constructed through a focused search---rather than an exhaustive systematic review---of recent work on scientific agents, AI for science, human--agent collaboration, tool-augmented workflows, peer review, and multi-agent scientific discovery. We use this coding to organize representative systems across puzzle-solving in normal science, anomaly evaluation, and paradigm-shift exploration. Full details on search sources, inclusion/exclusion criteria, screening, and coding decisions are provided in Appendix~\ref{corpuscoding}. Table~\ref{tab:hac_context_summary} summarizes, for each scientific context, the dominant coordination pattern that CSCW$\times$Kuhn identifies across these systems. The system-level coding is detailed in Appendix~\ref{app:system-coding} (Table~\ref{tab:hac_systems_modes}).

\begin{table*}[t]
\centering
\small
\setlength{\tabcolsep}{4pt}
\renewcommand{\arraystretch}{1.12}

\begin{tabularx}{\textwidth}{
>{\raggedright\arraybackslash}p{2.4cm}
>{\raggedright\arraybackslash}p{2.8cm}
>{\raggedright\arraybackslash}X
>{\raggedright\arraybackslash}p{2.8cm}
>{\raggedright\arraybackslash}p{2.8cm}
}
\toprule
\textbf{Scientific Context} &
\textbf{Representative System Clusters} &
\textbf{Coordination Focus} &
\textbf{Main Failure Risks} &
\textbf{Evaluation Implications}\\
\midrule

\textbf{Puzzle-solving in normal science}
&
Literature/evidence workflows; design and optimization; protocol execution; shared planning
&
\coordcell
{task decomposition, protocol execution, workflow tracking;}
{domain constraints, data formats, methods, success criteria;}
{intermediate outputs, checkpoints, errors, deviations}
&
Procedural drift; grounding substitution; hidden execution errors; weak long-horizon oversight
&
Audit traces; intermediate state checks; tool-use inspection
\\

\midrule

\textbf{Anomaly evaluation}
&
Hypothesis and novelty evaluation; LLM-as-reviewer; rebuttal and meta-review; institutional review
&
\coordcell
{review criteria, novelty checks, rebuttal and decision workflows;}
{paper--literature--benchmark--venue alignment;}
{uncertainty, disagreement, unsupported claims, confidence signals}
&
Evaluative flattening; shallow fluency; majority-opinion bias; weak adjudication of contested claims
&
Claim--evidence alignment checks; disagreement and judgment preservation
\\

\midrule

\textbf{Paradigm-shift exploration}
&
Cross-domain ideation; falsification and feedback loops; recursive AI scientist workflows; verification/search; infrastructure
&
\coordcell
{evolving goals, search strategies, validation plans;}
{problem--concept--domain--evidence alignment;}
{idea provenance, retrieved evidence, rankings, experimental outcomes}
&
Grounding substitution; proxy optimization; epistemic black-boxing; unsupported or non-testable ideas
&
Assumption tracing; provenance checks; uncertainty reports; expert critiques; downstream validation
\\

\bottomrule
\end{tabularx}

\caption{Summary of coordination patterns with corresponding evaluation implications across scientific contexts. We note that workflow stages are not exclusive to a single epistemic mode. The same stage may recur across scientific contexts with different coordination demands, and a single mode may span multiple workflow stages.}
\label{tab:hac_context_summary}
\end{table*}

\subsection{Puzzle-Solving in Normal Science}
\label{sec3.1}

\paragraph{Stable task frames.}
Puzzle-solving in normal science refers to work conducted within accepted goals, methods, and evaluation criteria \citep{kuhn2012}. In this context, the main coordination challenge is reliable execution. Articulation work concerns task decomposition, tool use, protocol following, and workflow integration. Common ground depends on shared assumptions about domain constraints, data formats, methods, and success criteria. Awareness requires visible intermediate outputs, status updates, errors, and checkpoints.

\paragraph{Observed collaboration pattern.}
Systems in this context mainly support bounded scientific workflows, including literature retrieval, evidence screening, molecular design, protocol execution, and shared planning \citep{wang2025foundation, ye2025drugassist, kon2025curie, feng2026cocoa}. In evidence-oriented workflows, citation interpretation can depend on broader document context and source-level information surrounding citations \citep{zhang2021tdmcfc,zhang2022native}. Citation-function modeling has also been advanced through hybrid data augmentation and broader deep-learning approaches \citep{zhang2023hybrid,zhang2025deepbibliometrics}, while recent survey work examines the growing intersection between LLMs and citation analysis \citep{zhang2023llmcitation}. Agents typically assist by decomposing tasks, expanding search, generating candidates, executing procedures, or maintaining workflow artifacts. Human involvement remains critical for checking constraints, validating intermediate results, and correcting deviations.

\paragraph{Evaluation gap.}
Current evaluations often emphasize task completion, output correctness, or benchmark performance. These measures are useful but incomplete. They may show that an agent completes a predefined task, while missing whether the system preserved the intended workflow, handled intermediate states correctly, and supported human oversight in long-horizon processes. Recent work on long-horizon agent memory likewise shows that downstream task success may obscure whether an agent retrieves the correct historical evidence, particularly under constrained context budgets \citep{li2026ocr}. These limitations create a risk of \textit{procedural drift}, in which an agent's actions deviate from the intended workflow, protocol, constraints, or task state, even when the final output appears successful.

\paragraph{Worked example: experiment design.}
AAAR-1.0 \citep{Lou2025AAAR} evaluates scientific tasks with well-defined input--output specifications (see Table~\ref{tab:benchmark-eval-mapping} for its coverage mapping). In its experiment design task, an LLM receives pre-experiment paper context and generates an experiment plan with motivations. It is reported that while some generated experiments can be innovative and diverse, many others are trivial, infeasible, or divergent from the original objective. The findings concern output quality rather than the intermediate generation process. As an illustrative scenario grounded in this benchmark design, a system could produce an experiment plan and receive a favorable score for its output. Yet violations of intermediate constraints or workflow deviations could remain unobserved.

\subsection{Anomaly Evaluation}
\label{sec3.2}

\paragraph{Contested scientific judgment.}
Anomaly evaluation refers to the assessment of uncertain findings, contested claims, or contributions whose significance is not yet settled. This includes novelty assessment, peer review, rebuttal analysis, critique generation, and meta-review. In this context, the coordination challenge shifts from reliable execution to evaluative alignment. Here, humans and agents must compare evidence, surface disagreement, and make judgment criteria explicit.

\paragraph{Observed collaboration pattern.}
Systems in this context often convert scientific judgment into structured workflows \citep{xiong2025truthhypo, lin2025schnovel, zhu-etal-2025-deepreview, chang2025treereview, wu2026rbtact}. They retrieve comparison evidence, decompose review criteria, identify unsupported claims, expose reviewer disagreement, or support rebuttal reasoning. We therefore view these agents as most appropriate for structuring judgment rather than replacing human evaluation.

\paragraph{Coordination demands.}
Articulation work involves decomposing scientific merit into criteria such as novelty, soundness, relevance, actionability, and evidential support. Common ground requires alignment among the paper, prior literature, benchmarks, venue expectations, reviewer concerns, and rebuttal evidence. Awareness necessitates visibility into uncertainty, disagreements, unsupported claims, confidence signals, and revisions before humans make decisions.

\paragraph{Evaluation gap.}
Existing evaluations often rely on prediction accuracy, agreement with human labels, review-quality rankings, groundedness, specificity, or actionability. These metrics do not fully capture whether a system helps humans preserve minority views, distinguish substantive novelty from superficial difference, or adjudicate evidential conflict. This limitation is consequential because AI-assisted review systems may reshape reviewer confidence, rebuttal behavior, and decision outcomes \citep{latona2025aireviewlottery, liang2024monitoring, Jin2024AgentReviewEP, zhao2026apres}. We characterize this risk as \textit{evaluative flattening}, in which contested scientific judgment is reduced to simplified scores, rankings, or majority opinions that obscure multidimensional criteria and disagreement.

\paragraph{Worked example: novelty assessment.}
SchNovel \citep{lin2025schnovel} evaluates whether LLMs can identify the more novel paper in a pair using titles, abstracts, and optional metadata, with performance measured by accuracy (Table~\ref{tab:benchmark-eval-mapping}). The findings show that novelty judgments are sensitive to factors unrelated to scholarly contribution. Models exhibit position bias, and adding institutional affiliation can produce a preference for papers associated with a top research university. This illustrates how a structured novelty judgment can be affected by contextual cues, even when the evaluation reduces the decision to a novelty label.

\subsection{Paradigm-Shift Exploration}
\label{sec3.3}

\paragraph{Unstable goals and criteria.}
Paradigm-shift exploration refers to upstream work that expands the conceptual search space, connects distant bodies of knowledge, and tests whether speculative directions deserve further investigation \citep{kuhn2012}. Unlike normal-science puzzle-solving, this context does not assume that goals, comparison spaces, or validation criteria are stable already.

\paragraph{Observed collaboration pattern.}
Systems in this context support cross-domain ideation, falsification loops, recursive research workflows, verification search, and scientific infrastructure \citep{wang2024scimon, baek2025researchagent, radensky2026scideator, gottweis2026coscientist, ghareeb2026robin, zhang2025internagent, lu2026endtoend, romeraparedes2024mathematical, shojaee2025llmsr}. In our framework, these systems are most credible when speculative generation is paired with evidence retrieval, critique, executable checks, provenance tracking, or expert revision.

\paragraph{Coordination demands.}
Within our framework, paradigm-shift exploration can impose particularly strong coordination demands among the tasks analyzed. Articulation work must stabilize evolving goals, search strategies, and validation plans. Common ground is fragile, as agents may introduce unfamiliar concepts, distant analogies, or weakly related literature. Awareness is essential because researchers must inspect the provenance of the idea, retrieved evidence, rankings, experimental results, and verification signals before deciding whether a direction is scientifically meaningful.

\paragraph{Evaluation gap.}
Current evaluation remains limited for paradigm-shift exploration. Criteria such as novelty, plausibility, or benchmark success do not establish whether a system supports durable conceptual reframing. The relevant question is not necessarily whether an agent can generate surprising ideas, but whether those ideas remain grounded, testable, interpretable, and useful under scrutiny from experts and follow-up validation.

\paragraph{Worked example: scientific ideation.}
\citet{si2025llms} compare LLM-generated research ideas with ideas from expert NLP researchers using blinded expert evaluation. LLM-generated ideas were rated significantly more novel than human ideas, but slightly weaker on feasibility. The authors identify failures in LLM self-evaluation along with limited diversity in generation. They further note that idea-level judgments alone do not establish whether these advantages translate into meaningful research outcomes, and propose executing the ideas as full projects for end-to-end evaluation. This illustrates how favorable novelty judgments can coexist with unresolved questions about feasibility and downstream scientific validity (Table~\ref{tab:benchmark-eval-mapping}).

\paragraph{Recurring failure modes.}
Over the three scientific contexts, we identify patterns of five coordination failure modes. The term \textit{failure risk} is used to denote the potential for a given failure mode to arise in a system or context. \textit{Procedural drift} occurs when collaboration deviates from the intended workflow, protocol, constraint set, or task state. \textit{Grounding substitution} occurs when plausibility, fluency, or apparent novelty replaces evidential support or domain validity. \textit{Evaluative flattening} occurs when contested scientific judgment is reduced to simplified scores, rankings, or majority opinions. \textit{Proxy optimization} occurs when systems optimize benchmark metrics while losing alignment with scientific goals. \textit{Epistemic black-boxing} occurs when provenance, assumptions, rationale, intermediate states, tool use, or responsibility become difficult to inspect or contest.

\section{Benchmarks and Evaluation Methods}
\label{sec4}

Current benchmarks provide useful measures of task performance but only partially evaluate human--agent collaboration in scientific innovation. As summarized in Table~\ref{tab:benchmark-eval-mapping}, most benchmarks focus on task- or output-level metrics (e.g., retrieval quality, answer correctness, hypothesis plausibility, review quality, code execution, end-to-end task completion). These metrics improve comparability but often under-measure whether humans and agents preserve workflow integrity, maintain evidential grounding, surface disagreement, and support accountable scientific judgment.

\subsection{Limits of Task-Centric Evaluation}

\paragraph{Workflow integrity.}
Many benchmarks decompose scientific work into standalone tasks with clear input--output formats. This design supports reproducible evaluation but can obscure whether an agent follows the intended workflow across intermediate steps. As illustrated in Section~\ref{sec3.1}, such benchmarks are suitable for measuring local task success but are less able to detect \textit{procedural drift}, where an agent deviates from protocols or constraints. Considering this, evaluation should include trace audits, intermediate state checks, tool-use inspection, and procedure-fidelity measures in addition to final accuracy.

\paragraph{Evidential grounding.}
Output-level verification does not necessarily establish scientific grounding. Generated hypotheses or reviews may appear plausible while remaining weakly supported or insufficiently tied to the underlying scientific claims \citep{xiong2025truthhypo, ou2025claimcheck}. CLAIMCHECK \citep{ou2025claimcheck} further shows that LLMs struggle to associate reviewer-identified weaknesses with the specific claims they dispute, and to generate specific, claim-grounded critiques from scratch. As an illustrative scenario, a generated claim may seem well supported due to relevant material being retrieved; however, the cited evidence may not directly support the claim or cannot be traced. This reflects the risk of \textit{grounding substitution}, where plausibility, fluency, or apparent novelty replaces evidence. Evaluation should test whether claims are supported by relevant sources, whether evidence is traceable, and whether generated conclusions remain consistent with the retrieved literature and data.

\paragraph{Scientific judgment.}
Some benchmarks isolate criteria that are easier to score than scientific merit as a whole. As discussed in Section~\ref{sec3.2}, SchNovel \citep{lin2025schnovel} advances the evaluation of scholarly novelty; however, novelty is one dimension of scientific judgment, not a proxy for it in full. In anomaly evaluation, this creates the risk of \textit{evaluative flattening}, where contested judgments are reduced to a single score. Evaluation should instead preserve tradeoffs among multiple dimensions of scientific value: novelty, validity, feasibility, significance, evidence quality, and disagreement.

\paragraph{Exploratory validity.}
Paradigm-shift exploration is especially difficult to evaluate because goals, assumptions, and validation criteria are often unstable. As illustrated in Section~\ref{sec3.3}, favorable novelty judgments do not necessarily establish feasibility or downstream scientific validity. Existing ideation benchmarks assess novelty, feasibility, diversity, and quality of ideas, yet generated ideas may require expert assessment and empirical validation before becoming actionable \citep{si2025llms, Ruan2026LiveIdeaBench}. Evaluation should therefore include assumption tracing, provenance checks, uncertainty reports, critiques from experts, and validation from long-horizon follow-ups.

\subsection{Toward Coordination-Aware Evaluation}

\paragraph{Evaluation targets.}
An evaluation that accounts for coordination should complement task metrics with four additional targets. First, it should assess \textit{process fidelity}: whether agents preserve protocols, constraints, intermediate states, and intended workflows. Second, it should assess \textit{claim--evidence alignment}: whether generated claims are supported by relevant, traceable, and domain-appropriate evidence. Third, it should assess \textit{judgment preservation}: whether systems maintain multiple evaluation criteria and make disagreement visible. Fourth, it should assess \textit{epistemic transparency}, addressing the risk of epistemic black-boxing, by testing whether users can inspect assumptions, provenance, uncertainty, tool use, and responsibility \citep{zhang2026adaqecg}.

\paragraph{Human oversight.}
Human oversight should vary with epistemic context. In puzzle-solving settings, humans should primarily engage in verification, such as checking workflow adherence, tool use, constraint satisfaction, and intermediate outputs. In anomaly evaluation, humans should focus on overseeing comparisons of evidence, reviewer disagreement, minority perspectives, and distinctions between substantive novelty and superficial difference. In paradigm-shift exploration, oversight should involve examining how candidate ideas were generated, what assumptions shaped these ideas, what evidence supports them, and what types of follow-up would make them testable.

\paragraph{Implications for benchmark design.}
Future benchmarks should move beyond final-output scoring to evaluate collaboration as a process. This requires in-depth traces, interpretable intermediate artifacts, explicit evidence links, human checkpoints for intervention, and longitudinal validation. Such evaluation does not replace task accuracy, but instead situates accuracy within the broader coordination demands of scientific work. Under CSCW$\times$Kuhn, benchmark completion is not sufficient evidence for success; the operative criterion is whether the system can support reliable coordination under the epistemic conditions of the scientific workflow.

\section{Discussion}
\label{sec5}

\paragraph{Implications for evaluation.}
CSCW$\times$Kuhn reframes evaluation from a question of task completion by agents to whether a collaborative system—consisting of humans and agents—supports the coordination demands of scientific work. We emphasize this distinction because the same benchmark score may reflect different levels of scientific reliability across various epistemic contexts. Evaluation should not stop at output correctness in puzzle-solving settings; it should also test whether the system preserves procedure fidelity, constraint satisfaction, tool-use validity, and consistency in intermediate states. With anomaly evaluation, the issue goes beyond whether an agent predicts a human label or generates a plausible review; evaluations in this context should assess whether agents help preserve disagreement, compare alternatives, expose evidential gaps, and distinguish substantive novelty from superficial difference. Evaluation requires even greater caution in paradigm-shift exploration. Novelty and plausibility are insufficient unless the system also supports provenance, feasibility assessment, falsifiability, and follow-up validation. Thus, coordination-aware evaluation complements task metrics by asking what kinds of scientific accountability remain visible to humans.

\paragraph{Implications for system design.}
This perspective cautions against interpreting current benchmark progress as support for fully autonomous scientific discovery, and suggests limits to strong claims about end-to-end automation. For instance, ScienceAgentBench \citep{Chen2025ScienceAgentBench} evaluates bounded workflow tasks and frames agents as assisting scientists with data processing and visualization, not substitutes for scientific judgment. In a related vein, benchmarks for hypothesis generation and novelty assessment show that plausible or original outputs still require grounding, expert interpretation, and downstream validation \citep{si2025llms, xiong2025truthhypo}. System design should therefore prioritize modular assistance, explicit human checkpoints, interpretable intermediate artifacts, and domain-specific validation interfaces. Agent outputs should be treated as candidate analyses, hypotheses, reviews, or drafts rather than finalized scientific conclusions. This is crucial in long-horizon workflows where small errors can accumulate and amplify into misleading scientific claims.

\paragraph{Implications for peer review and institutional governance.}
Human--agent collaboration also raises institutional questions when agents participate in evaluative settings such as peer review. Review and meta-review help determine which work receives credibility, attention, and publication opportunities. AgentReview \citep{Jin2024AgentReviewEP} shows that review outcomes are shaped by reviewer heterogeneity, latent variables, and privacy constraints, and that biased reviewers, in particular, can materially affect decisions. Related evidence suggests that AI-assisted reviewing can influence review scores and acceptance outcomes, notably for borderline submissions \citep{latona2025aireviewlottery}. Findings imply that the main governance issue is under what disclosure, monitoring, audit, and accountability regimes AI should be used. Without such safeguards, agentic judgments can shape scientific decisions without a clear record of how or by whom. Institutions therefore need policies that specify when AI assistance is permitted, how its use should be disclosed, how review evidence should be preserved, and how responsibility is assigned \citep{zhang2026adaqecg}.

\paragraph{Open challenges.}
Several challenges remain. First, coordination quality is difficult to benchmark: it is processual and often visible only through intermediate traces, so future benchmarks need richer logs, evidence links, and human-in-the-loop protocols. Second, long-horizon scientific value is hard to measure. Ideas that appear novel at the time of evaluation may later prove to be unproductive; thus, longitudinal studies are needed. Third, CSCW$\times$Kuhn should be empirically validated, testing whether researchers can reliably annotate its dimensions across real workflows. Finally, governance should be integrated into system development rather than treated as post-hoc.

\section{Conclusion}
\label{sec6}
This survey frames Human--Agent Collaboration in Scientific Innovation as an epistemically conditioned coordination process. We introduce CSCW$\times$Kuhn to connect scientific context, coordination demand, failure mode, and evaluation target across puzzle-solving, anomaly evaluation, and paradigm-shift exploration. Our synthesis suggests that current evaluations often under-measure workflow integrity, evidential grounding, disagreement preservation, transparency, and human oversight. We argue that assessing scientific agents requires looking past task performance to whether agents support accountable coordination across various epistemic conditions.

\newpage

\section*{Limitations}
\label{sec:limitations}
This survey has several limitations. First, the literature on agentic scientific systems is expanding rapidly, so our corpus cannot exhaustively cover all emerging systems, benchmarks, and deployment practices. Second, our coding relies on information reported in published papers; when system designs or evaluation procedures are underspecified, coordination demands and failure risks are interpreted from available evidence. Third, CSCW$\times$Kuhn codes the dominant epistemic condition of each system or workflow segment, while real scientific workflows may transition across puzzle-solving, anomaly evaluation, and paradigm-shift exploration. Future work can extend CSCW$\times$Kuhn through larger corpora, empirical annotation studies, and deployment-based validation.

\section*{Ethics Statement}
This paper is a survey of human--agent collaboration in scientific innovation and does not introduce new human-subject experiments or collect personal or sensitive data. We release a bibliographic corpus and analytic coding artifact derived from publicly available research papers and benchmarks. We acknowledge that human--agent scientific systems may raise ethical concerns related to accountability, transparency, overreliance on automated judgments, and the potential misuse of agentic systems in scientific evaluation. Accordingly, we emphasize that such systems should be designed and evaluated with human oversight, provenance tracking, uncertainty disclosure, and clear responsibility assignment.

\section*{Information About Use of AI Assistants}
AI assistants were used only for polishing language, correcting grammar, and improving the clarity and readability of the manuscript. They were not used to generate original scientific claims, conduct literature screening, make coding decisions, perform analysis, or produce conclusions without author verification. All content, references, interpretations, and final writing decisions were reviewed and approved by the authors, who take full responsibility for the manuscript.

\section*{Acknowledgments}
We would like to thank the anonymous reviewers and meta-reviewer for their constructive feedback, which helped us improve our presentation and clarity of this survey.

\bibliography{custom}

\appendix

\section{Survey Positioning}
\label{app:survey-positioning}
This appendix provides supplementary details for the survey scope, CSCW and Kuhnian conceptual grounding, operational coding rubric, corpus construction, analytic coding, and extended tables. Appendix~\ref{app:related-surveys} positions our work relative to existing surveys. Appendix~\ref{app:cscw-foundations} details the CSCW foundations of our coordination coding. Appendix~\ref{kuhnlen} clarifies how Kuhn's account of scientific change is used as an analytical lens. Appendix~\ref{app:coding-rubric} provides the operational rubric for coding epistemic context, coordination demand, failure modes, and evaluation targets. Appendix~\ref{corpuscoding} describes corpus construction and coding procedures. Appendix~\ref{app:system-coding} reports details of system-level coding used to support the synthesis in the main text. Appendix~\ref{app:supplementary-tables} provides additional tables for operational definitions, survey comparison, and benchmark mapping.

\subsection{Comparison with Existing Surveys}
\label{app:related-surveys}

To position CSCW$\times$Kuhn, we compare our survey with prior work on AI for science, agentic systems, multi-agent systems, and human--agent collaboration. These survey traditions provide important foundations, but differ in primary analytical focus. AI4Science surveys typically organize systems by scientific tasks or research pipelines. Agentic-system surveys focus on model architecture, planning, tool use, memory, and multi-agent coordination. Human--agent collaboration surveys provide stronger accounts of human roles, interaction, and agency distribution.

Our survey pursues a different analytical objective. Whereas prior surveys mainly organize systems by task, architecture, or application domain, we analyze human--agent scientific collaboration as an epistemically conditioned coordination process---connecting scientific context, coordination demand, human role, evaluation targets, and failure modes in a unified coding framework. We do not intend to reclassify prior surveys' systems, nor suggest that their conclusions are incorrect. We instead reinterpret representative systems through a coordination-aware lens to address the analytical coverage gap. Table~\ref{tab:survey_gap_comparison_full} summarizes our positioning.

\subsection{CSCW Foundations for Coordination Coding}
\label{app:cscw-foundations}

CSCW provides the conceptual foundation for the coordination dimensions used in CSCW$\times$Kuhn. We use CSCW as a theory of cooperative work that emphasizes the practical work required to make collaboration possible. Prior research in CSCW has shown that cooperative systems must support task execution, and also the organization of interdependent activities, the maintenance of shared understandings, and the visibility of ongoing work \citep{schmidt1992taking, schmidt1996coordination, dourish1992awareness, gutwin2002descriptive}.

This appendix clarifies how we translate CSCW concepts into coding dimensions. \textbf{Articulation work} refers to the organizing and repair work required to coordinate interdependent tasks, including sequencing, role assignment, handoffs, integration, and recovery from breakdowns \citep{strauss1985work, schmidt1992taking}. In human--agent scientific systems, this includes agent planning, tool selection, protocol execution, workflow decomposition, and human intervention points. \textbf{Common ground} refers to the extent to which humans and agents share sufficient understanding of goals, assumptions, constraints, evidence, methods, and task state for coordinated action \citep{clark1991grounding}. In scientific workflows, this includes alignment between research questions, retrieved literature, data, claims, benchmarks, and evaluation criteria. \textbf{Awareness} refers to the visibility of actions, intermediate states, errors, decisions, and consequences during collaborative work \citep{dourish1992awareness, gutwin2002descriptive}. In human--agent scientific systems, this includes logs, provenance, uncertainty signals, rationales, disagreement traces, and checkpoints that allow humans to monitor and contest agent behavior.

These three dimensions were selected because they are theoretically grounded in CSCW and operationally useful for analyzing agentic scientific workflows. The dimensions allow us to code for whether a system supports the coordination work needed for reliable scientific collaboration, rather than only if it produces a correct output. They also connect directly to the failure modes identified in the main text: procedural drift reflects failures of articulation work; grounding substitution reflects failures of common ground; evaluative flattening and epistemic black-boxing reflect failures of awareness, inspectability, and contestability.

\subsection{Kuhnian Lens for Scientific Contexts}
\label{kuhnlen}

We use Kuhn's account of scientific change as an analytical heuristic for distinguishing different epistemic conditions in scientific work \citep{kuhn2012}. In Kuhn's framework, scientific communities operate within paradigms that define legitimate problems, accepted methods, evidential standards, and criteria for successful explanation. Under stable paradigms, most work proceeds as normal science, where researchers solve puzzles within established assumptions. Scientific work may also involve anomalies (i.e., findings or claims that do not fit prevailing expectations). Some anomalies are resolved within the existing paradigm, whereas others motivate deeper conceptual reconsideration.

This distinction is useful for analyzing human--agent collaboration since different epistemic conditions impose different coordination demands. Puzzle-solving in normal science requires reliable execution, workflow tracking, and constraint management. Anomaly evaluation requires comparison among competing interpretations, careful treatment of evidence, and preservation of disagreement. Paradigm-shift exploration requires broader conceptual search, provenance over speculative ideas, and mechanisms for feasibility, falsifiability, and follow-up validation.

Accordingly, we operationalize three epistemic modes: \textit{puzzle-solving in normal science}, \textit{anomaly evaluation}, and \textit{paradigm-shift exploration}. These modes are not intended as exhaustive categories of all scientific activity, nor do they imply that real workflows remain fixed within a single mode. Instead, they provide a tractable coding scheme for analyzing how coordination demands change as scientific work moves from stable execution to contested judgment and exploratory reframing.

\section{Operational Coding Rubric}
\label{app:coding-rubric}

We operationalize CSCW$\times$Kuhn as a qualitative coding scheme that links epistemic context, coordination demand, failure mode, and evaluation target. The primary coding unit is the named system or artifact. When a paper presents clearly separable workflow components, those components inform the system-level coding, but each system receives one dominant epistemic-mode assignment for comparative analysis. For each system, we first identify the dominant epistemic context, then assess coordination support along three CSCW dimensions: articulation work ($\alpha$), common ground ($\gamma$), and awareness ($\omega$). We then identify potential failure risks, corresponding to the failure modes defined in the main text, and map them to evaluation targets.

\paragraph{Epistemic context.}
\textbf{Puzzle-solving in normal science} is assigned when the task has stable goals, established methods, and predefined success criteria, such as protocol execution, benchmarked prediction, evidence extraction, or constrained optimization. \textbf{Anomaly evaluation} is assigned when the task involves judging uncertain, contested, novel, or potentially problematic claims, such as peer review, novelty assessment, rebuttal analysis, critique, or claim verification. \textbf{Paradigm-shift exploration} is assigned when the task involves open-ended search for alternative hypotheses, cross-domain analogies, conceptual recombination, or long-horizon research directions where goals and evaluation criteria are not fully stable.

\paragraph{Coordination demand.}
We code each CSCW coordination dimension on an ordinal 0--3 scale. These scores are not treated as quantitative performance measures; rather, they provide a transparent way to compare what forms of coordination support are made available by different systems.

\begin{itemize}[leftmargin=*]
    \item \textbf{Articulation work ($\alpha$).}
    0 = no task decomposition or coordination structure;
    1 = fixed pipeline with predefined steps;
    2 = dynamic planning with explicit human checkpoints;
    3 = adaptive replanning across agents and humans, including handoffs, repair, and revision.

    \item \textbf{Common ground ($\gamma$).}
    0 = no shared state or explicit task context;
    1 = prompt-level task instruction only;
    2 = explicit shared memory, workspace, or task state;
    3 = shared assumptions, evidence, uncertainty, constraints, and role model are visible and maintained.

    \item \textbf{Awareness ($\omega$).}
    0 = final output only;
    1 = logs or execution traces are available;
    2 = intermediate states, provenance, and decision paths are visible;
    3 = humans can inspect, contest, intervene, and revise system actions or outputs.
\end{itemize}

\paragraph{Worked coding examples.}
To illustrate how the ordinal rubric applies in practice, let us walk through two examples of the coding process and the rationale behind the coordination demand scores. Consider the systems: LEADS \citep{wang2025foundation} and ResearchAgent \citep{baek2025researchagent}.

LEADS \citep{wang2025foundation} is coded at the system level as puzzle-solving in normal science with $(\alpha,\gamma,\omega)=(1,2,3)$. Its literature-mining workflow is decomposed into six predefined subtasks; thus \textbf{\textit{articulation work}} is coded as $\alpha=1$ \citep{wang2025foundation}. The reason why it is not coded as $\alpha=2$ or $3$ is because task order and structure are set in advance, and the paper does not describe any mechanism for dynamic task replanning. \textbf{\textit{Common ground}} is coded as $\gamma=2$, since screening and extraction operate in a structured task context, including PICO criteria, target-field definitions, eligibility assessments, and shared persistent artifacts (ranked spreadsheet) \citep{wang2025foundation}. That exceeds bare prompt-level instruction ($\gamma=1$), though no explicit role model or shared assumptions are maintained ($\gamma=3$). \textbf{\textit{Awareness}} is coded as $\omega=3$ because human experts can inspect criterion-level predictions and rationales, select studies independently of the AI ranking, and directly edit AI-extracted field values before finalizing them \citep{wang2025foundation}. This would not be a simple decision log ($\omega=1$) and exceeds $\omega=2$.

ResearchAgent \citep{baek2025researchagent} is coded at the system level as paradigm-shift exploration with $(\alpha,\gamma,\omega)=(1,2,2)$. Although its problem, method, and experiment-design outputs are iteratively revised by ReviewingAgents, the workflow remains predefined, and the paper does not report any inclusion of human checkpoints during generation, yielding $\alpha=1$ for \textbf{\textit{articulation work}} rather than $2$ (replanning is agent-only, not human-in-the-loop) or $0$ (a decomposed structure does exist) \citep{baek2025researchagent}. \textbf{\textit{Common ground}} is coded as $\gamma=2$ because the workflow maintains structured context across stages through prior idea components and rationales, retrieved literature, an entity-centric knowledge store, and shared evaluation criteria \citep{baek2025researchagent}. The workflow exceeds prompt-level instruction ($\gamma=1$) but lacks any persistent, explicit layer for uncertainty or assumptions required for $\gamma=3$. \textbf{\textit{Awareness}} is coded as $\omega=2$ because rationales, reviews, and ratings expose intermediate evaluative state \citep{baek2025researchagent}. This goes beyond simple logs or execution traces ($\omega=1$). However, in this system, human researchers participate primarily in post-hoc evaluation rather than in-loop intervention or revision required for $\omega=3$ \citep{baek2025researchagent}.

These ordinal scores are qualitative coding judgments about the coordination support exposed by the system design, not performance measures or claims about system quality.

\paragraph{Failure risk and evaluation target.}
Potential failure risks are identified by comparing the epistemic demands of a task with the coordination support provided by the system. \textbf{Procedural drift} is coded when a system performs multi-step work with limited task-state tracking, checkpointing, or replanning support. \textbf{Grounding substitution} is coded when outputs depend on evidence but provide weak provenance, unsupported claims, or insufficient source alignment. \textbf{Evaluative flattening} is coded when contested scientific judgment is reduced to narrow scores, rankings, or majority opinions. \textbf{Proxy optimization} is coded when a system is primarily optimized for benchmark metrics without evidence of downstream scientific validity. \textbf{Epistemic black-boxing} is coded when assumptions, intermediate reasoning, provenance, tool use, or responsibility boundaries are difficult to inspect or contest.

The corresponding evaluation targets are: workflow coordination for procedural drift, claim--evidence grounding for grounding substitution, multidimensional scientific judgment for evaluative flattening, downstream validation for proxy optimization, and inspectability or contestability for epistemic black-boxing.

\section{Corpus Construction and Analytic Coding}
\label{corpuscoding}

\subsection{Survey Protocol}
\label{app:survey-protocol}

Our goal was to construct a representative corpus of agentic systems, benchmarks, and evaluation studies relevant to human--agent collaboration in scientific innovation. The corpus was designed to support conceptual synthesis and framework development rather than to serve as an exhaustive systematic review. We therefore prioritized breadth across scientific workflows, agentic system designs, and evaluation settings.

\paragraph{Search sources.}
We searched ACL Anthology, arXiv, OpenReview, ACM Digital Library, IEEE Xplore, Google Scholar, and publisher websites for recent work on scientific agents, AI for science, LLM-based scientific workflows, multi-agent systems, and human--agent collaboration. We prioritized work published between 2024 and 2026 because this period captures the rapid emergence of LLM-based agents, tool-augmented systems, and multi-agent scientific workflows. Earlier work was included when it provided important conceptual grounding or representative system designs.

\paragraph{Search keywords.}
Search queries combined terms related to scientific workflows, language agents, collaboration, and evaluation. Representative queries included: \textit{LLM scientific discovery}, \textit{AI scientist}, \textit{scientific agents}, \textit{AI for science}, \textit{AI4Science}, \textit{human-agent collaboration}, \textit{human-AI collaboration}, \textit{human-AI collaborative hypothesis generation}, \textit{RAG scientific literature}, \textit{LLM-as-reviewer}, \textit{automated peer review}, \textit{scientific ideation LLM}, \textit{hypothesis generation LLM}, \textit{tool-use agents science}, \textit{tool-augmented LLM scientific workflow}, and \textit{multi-agent scientific discovery}.

\enlargethispage{\baselineskip}
\vspace{-4pt}

\paragraph{Screening process.}
Candidate papers were screened in three stages. First, titles and abstracts were screened for relevance to scientific research, discovery, evaluation, or scholarly workflows. Second, full papers were assessed to determine whether the system involved an agentic or LLM-based component relevant to planning, retrieval, critique, judgment, tool use, experimentation, writing, or evaluation, with continuous human input during execution not required for inclusion. Systems with substantial internal agent--agent coordination were included when scientific goals, constraints, validation standards, or accountability remained human-governed; these were treated as human--agent scientific collaboration at the broader sociotechnical level, with agent--agent coordination analyzed as an architectural mechanism within that workflow. Third, papers were selected for synthesis if they contributed analytical coverage for at least one epistemic mode or system cluster in our framework. 

\enlargethispage{\baselineskip}
\vspace{-4pt}

\begin{figure}[t]
    \centering
    \includegraphics[
        width=1.10\linewidth,
        trim=90 120 60 130,
        clip
    ]{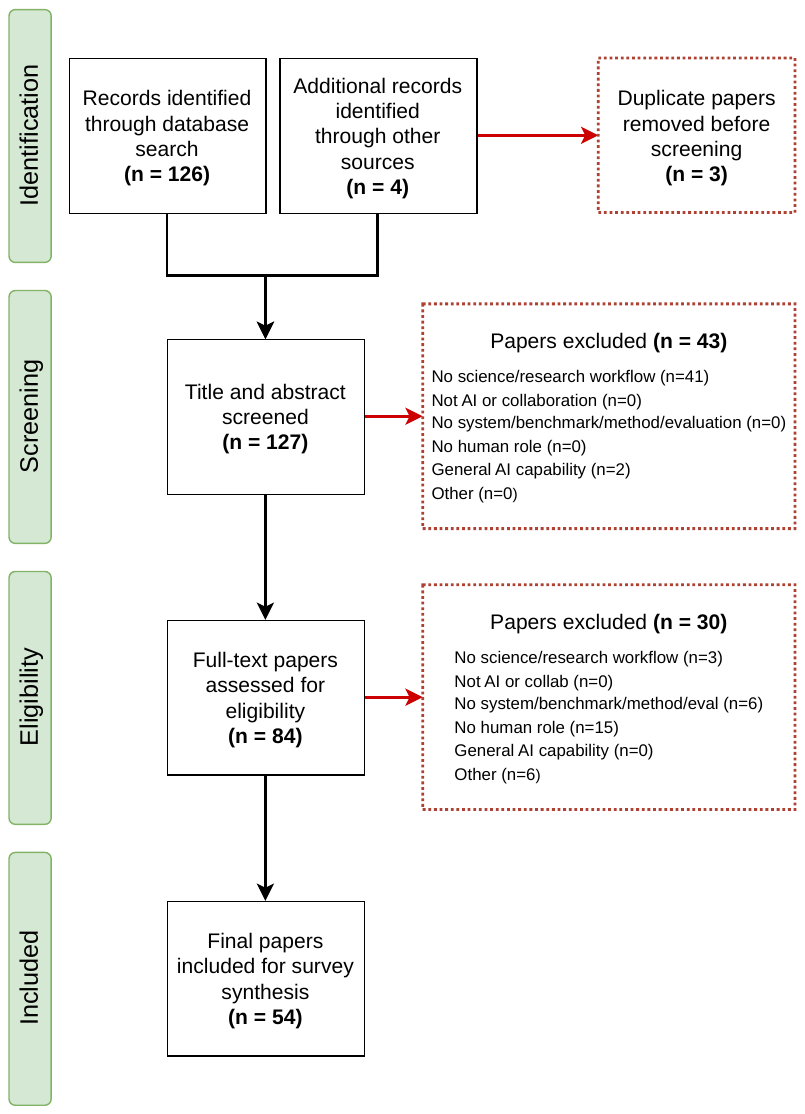}
    \caption{Flow diagram of our corpus selection.}
    \label{fig:prisma-flow}
    \vspace{-12pt}
\end{figure}

\paragraph{Screening results.}
We identified a total of 130 records. After removing 3 duplicate records, $127$ unique candidate papers were screened. During title and abstract screening, $43$ papers were excluded, leaving $84$ papers for full-text assessment. After full-text assessment, $54$ papers were included in the final synthesis, representing $59$ named systems and related artifacts. Figure~\ref{fig:prisma-flow} presents a flowchart of the survey protocol, inspired and adapted from the PRISMA framework \citep{moher2009prisma, page2021prisma} for transparent reporting. As this work is a focused conceptual survey and not a systematic review, we adopt its reporting structure selectively rather than strictly. The accompanying corpus and system-level coding artifact is
maintained at \url{https://github.com/jenlia024/CSCWxKuhn-artifact}.

\subsection{Analytic Coding}
\label{app:coding}

\paragraph{Inclusion criteria.}
We included papers that met at least two of the following criteria: 
(1) the system supports scientific research, scientific discovery, scholarly evaluation, or research-workflow automation; 
(2) the system uses LLMs, agents, retrieval-augmented generation, tool use, multi-agent coordination, review, critique, or ideation; 
(3) the system includes or implies a human role in task planning, judgment, oversight, revision, interpretation, or validation; and 
(4) the paper reports a concrete system, benchmark, method, or evaluation study relevant to scientific work.

\paragraph{Exclusion criteria.}
We excluded papers that focused only on general-purpose NLP tasks without a scientific workflow, domain-specific prediction without an agentic or collaborative component, purely theoretical discussions without a system or evaluation target, and applications where human involvement was absent, incidental, or not relevant to the scientific output. We also excluded papers whose relevance was limited to background AI capability rather than scientific collaboration, evaluation, or workflow coordination. Papers that met the inclusion criteria were removed if they also met any exclusion criterion.

\paragraph{Coding dimensions.}
Each included work was coded along four dimensions: \textit{epistemic mode}, \textit{system cluster}, \textit{coordination demand}, and \textit{potential failure risk}. The coding was designed to support the comparative synthesis in the main text rather than to produce mutually exclusive labels for all possible system properties.

\begin{itemize}[leftmargin=*]
    \item \textbf{Epistemic mode} identifies the dominant scientific context supported by the system: puzzle-solving in normal science, anomaly evaluation, or paradigm-shift exploration. When a system could fit multiple modes, we assigned the mode corresponding to its primary workflow role and evaluation target.

    \item \textbf{System cluster} captures the dominant workflow function of the system, such as literature and evidence workflows, design and optimization, protocol execution, shared orchestration, novelty evaluation, peer review, rebuttal support, cross-domain ideation, falsification, recursive AI-scientist workflows, verification/search, or scientific infrastructure.

    \item \textbf{Coordination demand} records how the system supports the three CSCW dimensions used in our framework: articulation work $(\alpha)$, common ground $(\gamma)$, and awareness $(\omega)$. Articulation work concerns task organization and workflow integration; common ground concerns shared assumptions, goals, evidence, and constraints; awareness concerns the visibility of actions, intermediate states, rationales, errors, and outcomes.

    \item \textbf{Potential failure risk} identifies the main coordination vulnerability implied by the system design, task setting, interface, or evaluation setup. These risks were used to synthesize recurring failure modes across systems.
\end{itemize}

\paragraph{Failure-mode synthesis.}
After coding individual systems, we compared failure risks across epistemic modes and system clusters. We synthesized five recurring coordination failure modes. \textbf{Procedural drift} refers to deviations from intended workflows, protocols, constraints, or task states. \textbf{Grounding substitution} occurs when fluency, plausibility, or apparent novelty substitutes for evidential support or domain validity. \textbf{Evaluative flattening} occurs when contested scientific judgment is reduced to simplified scores, rankings, criteria, or majority opinions. \textbf{Proxy optimization} occurs when systems optimize benchmark metrics or local evaluation signals while losing alignment with broader scientific goals. \textbf{Epistemic black-boxing} occurs when provenance, assumptions, intermediate reasoning, tool use, or responsibility become difficult for humans to inspect, contest, or revise.

\paragraph{Use of coding in the survey.}
The coding was used in two ways. First, it supported the system-level synthesis in Table~\ref{tab:hac_systems_modes}, where representative systems are grouped by epistemic mode, system cluster, coordination demand, and failure risk. Second, it supported the benchmark analysis in Table~\ref{tab:benchmark-eval-mapping}, where evaluation studies are compared according to their coverage of workflow coordination, human oversight, scientific judgment, and downstream validation. This procedure allows the survey to move beyond a list of systems and instead analyze how coordination requirements change across scientific contexts.

\paragraph{Reliability and limitations.}
The current coding should be interpreted as analytic coding for conceptual synthesis, not as a finalized systematic-review annotation. Coding decisions were checked against the reported system design, workflow role, and evaluation setup of each paper. Because many systems span multiple scientific stages, epistemic modes are not treated as mutually exclusive categories. We assigned the dominant mode to support comparison, while acknowledging that real scientific workflows may move from routine puzzle-solving to anomaly evaluation or exploratory reframing. Future work can extend this codebook through independent annotators, inter-annotator agreement, larger corpora, and deployment-based validation.

\section{System-Level Coding}
\label{app:system-coding}

Table~\ref{tab:hac_systems_modes} summarizes the system-level coding used to construct the synthesis in the main text. For each epistemic mode, we group representative systems into workflow-oriented clusters and identify their primary coordination demands and potential failure risks. Icons indicate artifact type, while the coordination-demand column reports articulation work \((\alpha)\), common ground \((\gamma)\), and awareness \((\omega)\). Ordinal coding and metadata are provided in the machine-readable artifact linked in Appendix~\ref{corpuscoding}.

\section{Supplementary Tables}
\label{app:supplementary-tables}

This section provides supplementary tables that support the main analysis. Table~\ref{tab:hac_systems_modes} provides the full system-level coding of representative human--agent scientific collaboration systems, grouped by epistemic mode and workflow cluster. Table~\ref{tab:survey_gap_comparison_full} compares existing survey traditions with our survey. Table~\ref{tab:operational-definitions} defines the main terms used throughout the paper. Table~\ref{tab:benchmark-eval-mapping} maps benchmark and evaluation studies by scientific domain, target task, metric, and coverage gap. The coding procedure behind these tables is described in Appendix~\ref{app:coding-rubric}.

\begin{table*}[!p]
\centering
\tiny
\setlength{\tabcolsep}{3pt}
\renewcommand{\arraystretch}{1.15}

\providecommand{\masicon}{%
  \raisebox{-0.15em}{\includegraphics[height=1.05em]{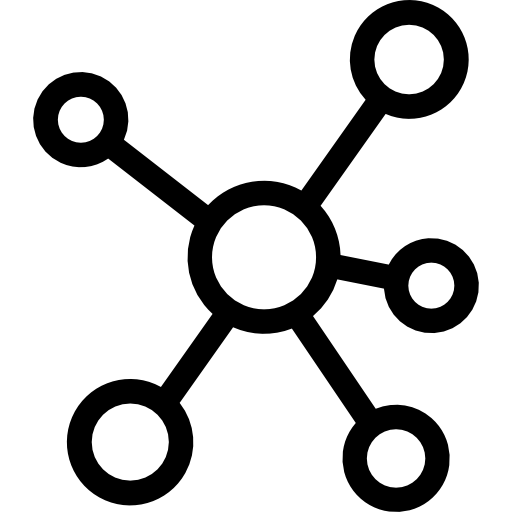}}%
}
\providecommand{\methodicon}{\includegraphics[height=1.1em]{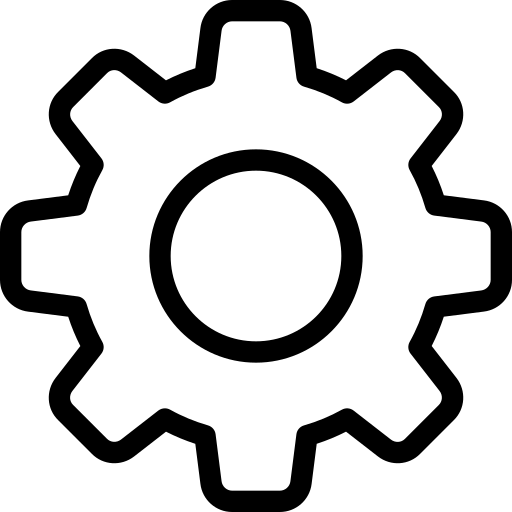}}
\providecommand{\interfaceicon}{\includegraphics[height=1.1em]{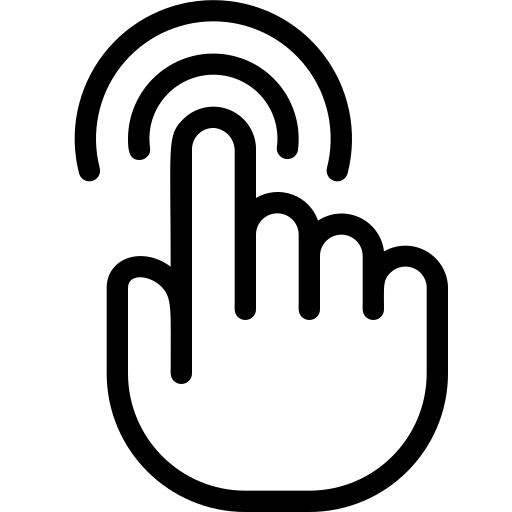}}
\providecommand{\infraicon}{\includegraphics[height=1.1em]{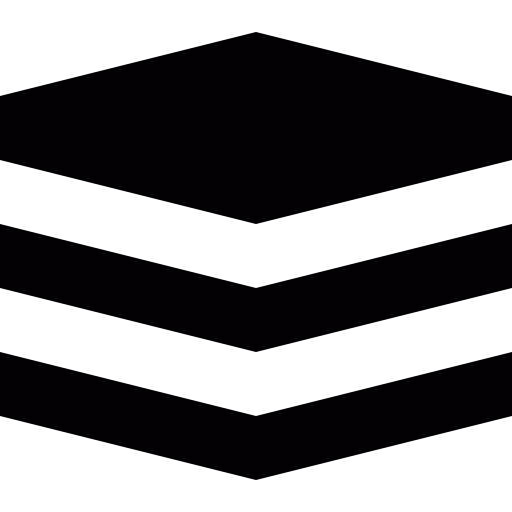}}
\providecommand{\benchicon}{\includegraphics[height=1.1em]{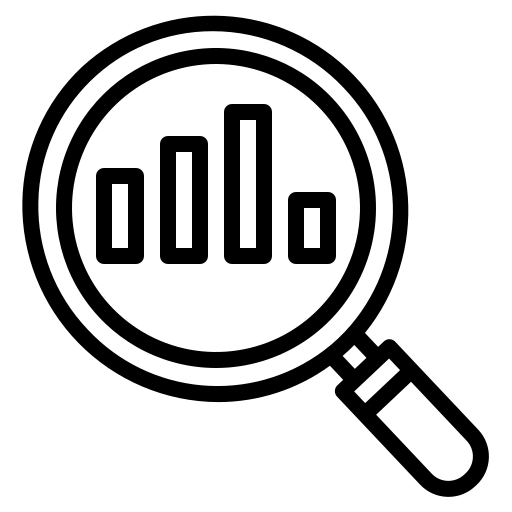}}
\providecommand{\agenticon}{\includegraphics[height=1.1em]{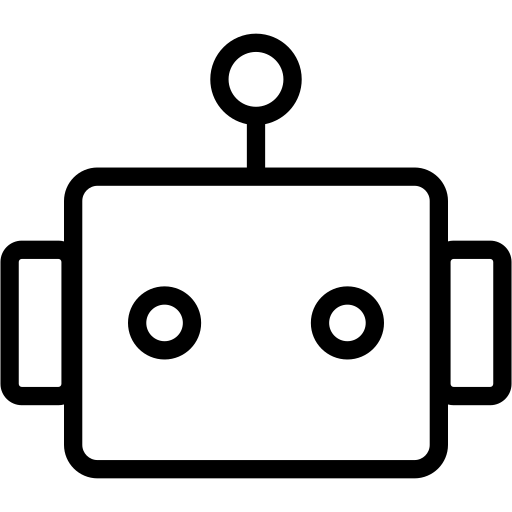}}
\providecommand{\frameworkicon}{\includegraphics[height=1.1em]{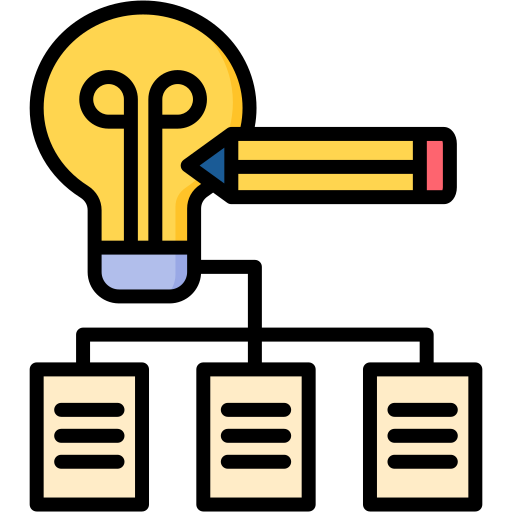}}
\providecommand{\modelicon}{\includegraphics[height=1.1em]{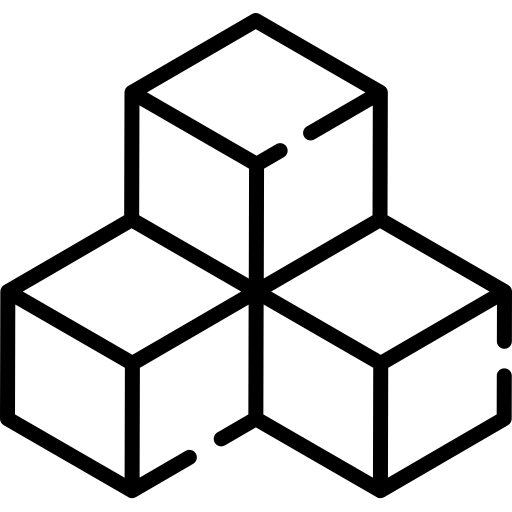}}

\begin{adjustbox}{max width=\textwidth, max totalheight=1\textheight, keepaspectratio}
\begin{tabularx}{\textwidth}{
>{\raggedright\arraybackslash}p{1.2cm}
>{\raggedright\arraybackslash}p{4.5cm}
>{\raggedright\arraybackslash}X
>{\raggedright\arraybackslash}p{2.5cm}
}
\toprule
\textbf{System Cluster} &
\textbf{Representative Systems} &
\textbf{Coordination Demand} &
\textbf{Potential Failure Risks} \\
\midrule

\multicolumn{4}{c}{\textbf{Puzzle-Solving in Normal Science}} \\
\midrule

Literature \& evidence workflows
&
[\modelicon] \textbf{LEADS} \citep{wang2025foundation}
&
\begin{minipage}[t]{\linewidth}
\setlength{\fboxsep}{2pt}
\colorbox{gray!15}{\parbox{\dimexpr\linewidth-2\fboxsep\relax}{\bm{$\alpha$}: subtask decomposition}}
\colorbox{white}{\parbox{\dimexpr\linewidth-2\fboxsep\relax}{\bm{$\gamma$}: PICO, eligibility, field alignment}}
\colorbox{gray!15}{\parbox{\dimexpr\linewidth-2\fboxsep\relax}{\bm{$\omega$}: ranked candidates, rationales, extraction outputs for expert oversight}}
\end{minipage}
&
Missed studies; borderline-case errors; extraction errors
\\

\midrule

Design \& optimization agents
&
[\frameworkicon] \textbf{ChatDrug} \citep{liu2023chatdrug}\newline
[\modelicon] \textbf{DrugAssist} \citep{ye2025drugassist}\newline
[\modelicon] \textbf{InstructBioMol} \citep{zhuang2025instructbiomol}\newline
[\frameworkicon, \masicon] \textbf{ChatBattery} \citep{liu2025chatbattery}\newline
[\frameworkicon, \modelicon] \textbf{MetaScientist} \citep{qi2025metascientist}
&
\begin{minipage}[t]{\linewidth}
\setlength{\fboxsep}{2pt}
\colorbox{gray!15}{\parbox{\dimexpr\linewidth-2\fboxsep\relax}{\bm{$\alpha$}: generation, editing, optimization subtasks}}
\colorbox{white}{\parbox{\dimexpr\linewidth-2\fboxsep\relax}{\bm{$\gamma$}: property--format--constraint alignment}}
\colorbox{gray!15}{\parbox{\dimexpr\linewidth-2\fboxsep\relax}{\bm{$\omega$}: candidates, feedback, explanations, validation scores}}
\end{minipage}
&
Unsafe, uninterpretable, or infeasible candidate molecules
\\

\midrule

Execution \& protocol systems
&
[\frameworkicon, \masicon] \textbf{Curie} \citep{kon2025curie}\newline
[\modelicon] \textbf{CycleResearcher} \citep{Weng2025CycleResearcher}\newline
[\masicon] \textbf{BioResearcher} \citep{luo2025bioresearcher}\newline
[\masicon, \interfaceicon] \textbf{LabOS} \citep{cong2025labos}
&
\begin{minipage}[t]{\linewidth}
\setlength{\fboxsep}{2pt}
\colorbox{gray!15}{\parbox{\dimexpr\linewidth-2\fboxsep\relax}{\bm{$\alpha$}: protocol planning, code execution, documentation}}
\colorbox{white}{\parbox{\dimexpr\linewidth-2\fboxsep\relax}{\bm{$\gamma$}: method--data--variable, reproducibility alignment}}
\colorbox{gray!15}{\parbox{\dimexpr\linewidth-2\fboxsep\relax}{\bm{$\omega$}: step state, quality checks, deviations, errors, results}}
\end{minipage}
&
Protocol failures; variable, dataset, or deviation mishandling
\\

\midrule

Shared planning \& orchestration systems
&
[\agenticon, \frameworkicon] \textbf{DrugAgent} \citep{liu2025drugagent}\newline
[\frameworkicon, \infraicon] \textbf{Agent Laboratory} \citep{schmidgall-etal-2025-agent}\newline
[\masicon] \textbf{MLZero} \citep{fang2025mlzero}\newline
[\frameworkicon, \masicon] \textbf{freephdlabor} \citep{li2025freephdlabor}\newline
\frameworkicon, \masicon] \textbf{data-to-paper} \citep{ifargan2025autonomous}\newline
[\masicon] \textbf{SciSciGPT} \citep{shao2026sciscigpt}\newline
[\frameworkicon, \masicon] \textbf{PaperOrchestra} \citep{song2026paperorchestra}\newline
[\interfaceicon] \textbf{Cocoa} \citep{feng2026cocoa}
&
\begin{minipage}[t]{\linewidth}
\setlength{\fboxsep}{2pt}
\colorbox{gray!15}{\parbox{\dimexpr\linewidth-2\fboxsep\relax}{\bm{$\alpha$}: plan--subtask--role--tool orchestration, review checkpoints}}
\colorbox{white}{\parbox{\dimexpr\linewidth-2\fboxsep\relax}{\bm{$\gamma$}: goal--artifact--constraint--workspace alignment}}
\colorbox{gray!15}{\parbox{\dimexpr\linewidth-2\fboxsep\relax}{\bm{$\omega$}: plans, reports, memory, feedback, intervention points}}
\end{minipage}
&
Context loss; role misalignment; degraded intermediate information
\\

\addlinespace
\toprule
\multicolumn{4}{c}{\textbf{Anomaly Evaluation}} \\
\midrule

Hypothesis generation \& novelty evaluation
&
[\benchicon, \methodicon] \textbf{SchNovel} / \textbf{RAG-Novelty} \citep{lin2025schnovel}\newline
[\benchicon, \frameworkicon] \textbf{TruthHypo} / \textbf{KnowHD} \citep{xiong2025truthhypo}\newline
[\modelicon, \methodicon] \textbf{Knowledge Interaction Model / Fusion module} \citep{wu2025novelty}\newline
[\modelicon] \textbf{HypER} \citep{vasu2025hyper}\newline
[\modelicon] \textbf{MoRI} \citep{gu2026mori}
&
\begin{minipage}[t]{\linewidth}
\setlength{\fboxsep}{2pt}
\colorbox{gray!15}{\parbox{\dimexpr\linewidth-2\fboxsep\relax}{\bm{$\alpha$}: novelty, truthfulness, grounding checks}}
\colorbox{white}{\parbox{\dimexpr\linewidth-2\fboxsep\relax}{\bm{$\gamma$}: review--metadata--literature alignment, knowledge graphs}}
\colorbox{gray!15}{\parbox{\dimexpr\linewidth-2\fboxsep\relax}{\bm{$\omega$}: scores, rationales, retrieved evidence, unsupported claims}}
\end{minipage}
&
Conflation of perceived novelty and plausibility with grounded substance
\\

\midrule

LLM-as-reviewer \& critique systems
&
[\frameworkicon, \benchicon] \textbf{DeepReview} /
[\modelicon] \textbf{DeepReviewer-14B} \citep{zhu-etal-2025-deepreview}\newline
[\modelicon] \textbf{CycleReviewer} \citep{Weng2025CycleResearcher}\newline
[\frameworkicon] \textbf{TreeReview} \citep{chang2025treereview}\newline
[\frameworkicon, \masicon] \textbf{MLR-Copilot} \citep{li2025mlrcopilot}\newline
[\frameworkicon] \textbf{ScholarPeer} \citep{goyal2026scholarpeer}
&
\begin{minipage}[t]{\linewidth}
\setlength{\fboxsep}{2pt}
\colorbox{gray!15}{\parbox{\dimexpr\linewidth-2\fboxsep\relax}{\bm{$\alpha$}: novelty, soundness, baseline checks}}
\colorbox{white}{\parbox{\dimexpr\linewidth-2\fboxsep\relax}{\bm{$\gamma$}: venue--literature--benchmark alignment}}
\colorbox{gray!15}{\parbox{\dimexpr\linewidth-2\fboxsep\relax}{\bm{$\omega$}: evidence-cited answers, retrieved literature, question trees, verification logs, confidence signals}}
\end{minipage}
&
Shallow fluency; missed flaws or baselines; hallucinated or unusable feedback
\\

\midrule

Rebuttal \& meta-review systems
&
[\benchicon] \textbf{ReviewCritique} \citep{du2024llms}\newline
[\frameworkicon] \textbf{CAF} \citep{chen2025cognitivealignment}\newline
[\modelicon, \benchicon, \methodicon] \textbf{RbtAct} \citep{wu2026rbtact}
&
\begin{minipage}[t]{\linewidth}
\setlength{\fboxsep}{2pt}
\colorbox{gray!15}{\parbox{\dimexpr\linewidth-2\fboxsep\relax}{\bm{$\alpha$}: deficiency, conflict, actionable feedback checks}}
\colorbox{white}{\parbox{\dimexpr\linewidth-2\fboxsep\relax}{\bm{$\gamma$}: paper--review--rebuttal--decision alignment}}
\colorbox{gray!15}{\parbox{\dimexpr\linewidth-2\fboxsep\relax}{\bm{$\omega$}: deficient review segments, conflict maps, rebuttal uptake, evidence signals}}
\end{minipage}
&
Majority-opinion bias; overreliance on reviews; weak adjudication
\\

\midrule

Institutional review systems
&
[\methodicon] \textbf{Distributional GPT quantification} \citep{liang2024monitoring}\newline
[\masicon, \frameworkicon] \textbf{AgentReview} \citep{Jin2024AgentReviewEP}\newline
[\frameworkicon] \textbf{APRES} \citep{zhao2026apres}
&
\begin{minipage}[t]{\linewidth}
\setlength{\fboxsep}{2pt}
\colorbox{gray!15}{\parbox{\dimexpr\linewidth-2\fboxsep\relax}{\bm{$\alpha$}: revision, review simulation, decision modeling, AI-use monitoring}}
\colorbox{white}{\parbox{\dimexpr\linewidth-2\fboxsep\relax}{\bm{$\gamma$}: rubric--reviewer role--rebuttal--decision alignment}}
\colorbox{gray!15}{\parbox{\dimexpr\linewidth-2\fboxsep\relax}{\bm{$\omega$}: revision comparisons, simulated review traces, decision shifts, AI-use prevalence}}
\end{minipage}
&
Black-box review or revision outputs; simplified evaluative judgments
\\

\addlinespace
\toprule
\multicolumn{4}{c}{\textbf{Paradigm-Shift Exploration}} \\
\midrule

Cross-domain ideation systems
&
[\frameworkicon] \textbf{SciMON} \citep{wang2024scimon}\newline
[\frameworkicon] \textbf{SciMuse} \citep{gu2025scimuse}\newline
[\masicon] \textbf{VIRSCI} \citep{su-etal-2025-many}\newline
[\frameworkicon, \masicon] \textbf{ResearchAgent} \citep{baek2025researchagent}\newline
[\frameworkicon, \methodicon] \textbf{CodeScientist} \citep{jansen2025codescientist}\newline
[\interfaceicon, \frameworkicon] \textbf{Scideator} \citep{radensky2026scideator}
&
\begin{minipage}[t]{\linewidth}
\setlength{\fboxsep}{2pt}
\colorbox{gray!15}{\parbox{\dimexpr\linewidth-2\fboxsep\relax}{\bm{$\alpha$}: literature retrieval, facet extraction, cross-domain recombination, novelty checking}}
\colorbox{white}{\parbox{\dimexpr\linewidth-2\fboxsep\relax}{\bm{$\gamma$}: research problem--concept--facet--domain alignment}}
\colorbox{gray!15}{\parbox{\dimexpr\linewidth-2\fboxsep\relax}{\bm{$\omega$}: inspirations, provenance, novelty signals, feedback, revision paths}}
\end{minipage}
&
Weakly grounded ideas; misleading analogies; vague research directions
\\

\midrule

Falsification \& feedback systems
&
[\masicon] \textbf{Baby-AIGS} \citep{liu2024aigs}\newline
[\masicon, \frameworkicon] \textbf{InternAgent} \citep{zhang2025internagent}\newline
[\masicon] \textbf{Kosmos} \citep{mitchener2025kosmos}\newline
[\masicon] \textbf{Robin} \citep{ghareeb2026robin}\newline
[\masicon, \frameworkicon] \textbf{AI Co-Scientist} \citep{gottweis2026coscientist}\newline
[\agenticon] \textbf{CellVoyager} \citep{alber2026cellvoyager}
&
\begin{minipage}[t]{\linewidth}
\setlength{\fboxsep}{2pt}
\colorbox{gray!15}{\parbox{\dimexpr\linewidth-2\fboxsep\relax}{\bm{$\alpha$}: hypothesis generation, critique, validation, data analysis, revision loops}}
\colorbox{white}{\parbox{\dimexpr\linewidth-2\fboxsep\relax}{\bm{$\gamma$}: objective--evidence--data/code--claim alignment}}
\colorbox{gray!15}{\parbox{\dimexpr\linewidth-2\fboxsep\relax}{\bm{$\omega$}: falsification/review agents, rankings, experiment results, citations, feedback}}
\end{minipage}
&
Amplified weak hypotheses; proxy overfitting; unsupported long-horizon claims
\\

\midrule

Recursive AI scientist workflows
&
[\agenticon, \frameworkicon] \textbf{The AI Scientist-v2} \citep{yamada2025aiscientistv2}\newline
[\frameworkicon, \masicon] \textbf{AI-Researcher} \citep{tang2025airesearcher}\newline
[\frameworkicon, \masicon] \textbf{DeepScientist} \citep{weng2026deepscientist}\newline
[\agenticon, \frameworkicon] \textbf{The AI Scientist} \citep{lu2026endtoend}
&
\begin{minipage}[t]{\linewidth}
\setlength{\fboxsep}{2pt}
\colorbox{gray!15}{\parbox{\dimexpr\linewidth-2\fboxsep\relax}{\bm{$\alpha$}: ideation, code, experimentation, writing, result analysis, recursive refinement}}
\colorbox{white}{\parbox{\dimexpr\linewidth-2\fboxsep\relax}{\bm{$\gamma$}: research goal--hypothesis--result--manuscript alignment}}
\colorbox{gray!15}{\parbox{\dimexpr\linewidth-2\fboxsep\relax}{\bm{$\omega$}: experiment logs, errors, plots, reviews, manuscript drafts, feedback}}
\end{minipage}
&
Ungrounded ideas; recursive errors; result misinterpretation; hidden methodological gaps
\\

\midrule

Verification \& search systems
&
[\methodicon, \frameworkicon] \textbf{FunSearch} \citep{romeraparedes2024mathematical}\newline
[\frameworkicon] \textbf{LLM-SR} \citep{shojaee2025llmsr}\newline
[\agenticon, \frameworkicon] \textbf{AlphaEvolve} \citep{novikov2025alphaevolve}
&
\begin{minipage}[t]{\linewidth}
\setlength{\fboxsep}{2pt}
\colorbox{gray!15}{\parbox{\dimexpr\linewidth-2\fboxsep\relax}{\bm{$\alpha$}: program/equation search and evaluation}}
\colorbox{white}{\parbox{\dimexpr\linewidth-2\fboxsep\relax}{\bm{$\gamma$}: problem spec--evaluator--code alignment}}
\colorbox{gray!15}{\parbox{\dimexpr\linewidth-2\fboxsep\relax}{\bm{$\omega$}: scores, evaluator outputs, program databases, best candidates}}
\end{minipage}
&
Proxy-score overfitting; high-scoring but weakly interpretable outputs
\\

\midrule

Infrastructure \& agent ecosystems
&
[\infraicon] \textbf{ToolUniverse} \citep{gao2025tooluniverse}\newline
[\infraicon, \frameworkicon] \textbf{OmniScientist} \citep{shao2025omniscientist}
&
\begin{minipage}[t]{\linewidth}
\setlength{\fboxsep}{2pt}
\colorbox{gray!15}{\parbox{\dimexpr\linewidth-2\fboxsep\relax}{\bm{$\alpha$}: tool, protocol, and provenance workflows}}
\colorbox{white}{\parbox{\dimexpr\linewidth-2\fboxsep\relax}{\bm{$\gamma$}: tool--data--knowledge-graph--norm alignment}}
\colorbox{gray!15}{\parbox{\dimexpr\linewidth-2\fboxsep\relax}{\bm{$\omega$}: traces, outputs, provenance, review signals, rankings}}
\end{minipage}
&
Hidden tool misuse; provenance gaps; weak governance; misaligned collaboration
\\

\bottomrule
\end{tabularx}
\end{adjustbox}

\vspace{0.5em}
\caption{Full system-level coding of representative human--agent scientific collaboration systems. Systems are grouped by epistemic mode and workflow cluster. Icons indicate artifact type: \modelicon{} model, \frameworkicon{} framework, \agenticon{} single-agent or agentic system, \masicon{} multi-agent system, \benchicon{} benchmark, \methodicon{} method, \interfaceicon{} interface, and \infraicon{} infrastructure. The coordination-demand column summarizes articulation work \((\alpha)\), common ground \((\gamma)\), and awareness \((\omega)\).}
\label{tab:hac_systems_modes}
\end{table*}

\begin{table*}[t]
\centering
\scriptsize
\setlength{\tabcolsep}{4pt}
\renewcommand{\arraystretch}{1.22}

\begin{tabularx}{\textwidth}{
    p{0.12\textwidth}
    p{0.15\textwidth}
    p{0.21\textwidth}
    X
    p{0.11\textwidth}
}
\toprule
\textbf{Survey Theme} &
\textbf{Survey Papers} &
\textbf{Scientific Epistemic Context} &
\textbf{Human Role Framing} &
\textbf{Explicit Failure Modes} \\
\midrule

\rowcolor{gray!10}
\cellcolor{white}
& \citet{Luo2025LLM4SR}
& \cmark\ Addressed
& \partialmark\ Addressed for peer review/writing; no human-role taxonomy
& \xmark\ \textcolor{red}{Absent} \\

\raggedright \textbf{AI4Science}
& \citet{chen2025ai4research}
& \xmark\ Not addressed; focuses on tasks and workflows
& \cmark\ Addressed
& \xmark\ \textcolor{red}{Absent} \\

\rowcolor{gray!10}
\cellcolor{white}
& \citet{wei2025agentic}
& \cmark\ Addressed
& \partialmark\ Discusses human roles in relation to agents; no human-role taxonomy
& \xmark\ \textcolor{red}{Absent} \\

& \citet{Zheng2025Automation}
& \cmark\ Addressed
& \cmark\ Addressed
& \xmark\ \textcolor{red}{Absent} \\

\rowcolor{gray!10}
\cellcolor{white}
& \citet{Tie2026AIScientists}
& \cmark\ Addressed
& \partialmark\ Discusses human roles in relation to agents; no human-role taxonomy
& \xmark\ \textcolor{red}{Absent} \\

\midrule

& \citet{wang2024autonomousagentsurvey}
& \partialmark\ Not central; general application focus
& \xmark\ Not addressed in scientific epistemic contexts
& \xmark\ \textcolor{red}{Absent} \\

\rowcolor{gray!10}
\cellcolor{white}\raggedright \textbf{Agentic System}
& \citet{guo2024multiagentsurvey}
& \partialmark\ Not central; general application focus
& \partialmark\ Weakly addressed; no human-role taxonomy
& \xmark\ \textcolor{red}{Absent} \\

& \citet{tran2025multiagentcollaboration}
& \xmark\ Not addressed
& \partialmark\ Weakly addressed; no human-role taxonomy
& \xmark\ \textcolor{red}{Absent} \\

\rowcolor{gray!10}
\cellcolor{white}
& \citet{xu2025toollearningsurvey}
& \xmark\ Not addressed
& \partialmark\ Weakly addressed; no human-role taxonomy
& \xmark\ \textcolor{red}{Absent} \\

& \citet{jiang2025adaptation}
& \cmark\ Addressed
& \xmark\ Not addressed
& \xmark\ \textcolor{red}{Absent} \\

\midrule

\rowcolor{gray!10}
\cellcolor{white}
& \citet{zahedi2021symbiosis}
& \xmark\ Not addressed
& \cmark\ Addressed
& \xmark\ \absent \\

\cellcolor{white}\raggedright \textbf{Human--Agent\\Collaboration}
& \citet{vaccaro2024combinations}
& \xmark\ Not addressed; general application focus
& \partialmark\ Weakly addressed; no human-role taxonomy
& \xmark\ \absent \\

\rowcolor{gray!10}
\cellcolor{white}
& \citet{gomez2024notcollaborative}
& \xmark\ Not addressed; healthcare application
& \cmark\ Broadly addressed in decision-making settings
& \xmark\ \absent \\

\cellcolor{white}
& \citet{vats2025lfm}
& \xmark\ Not addressed; healthcare application
& \cmark\ Addressed
& \notexplicit \\

\rowcolor{gray!10}
\cellcolor{white}
& \citet{zhang2025agency}
& \xmark\ Not addressed
& \cmark\ Addressed
& \xmark\ \absent \\

\cellcolor{white}
& \citet{shi2026scientificdiscovery}
& \cmark\ Addressed
& \cmark\ Addressed
& \notexplicit \\

\rowcolor{gray!10}
\cellcolor{white}
& \citet{Zou2026LLMHAS}
& \xmark\ Not addressed
& \cmark\ Addressed
& \notexplicit \\

\midrule

\rowcolor{gray!10}
\cellcolor{white}\raggedright \textbf{This Survey}
& \textbf{Ours}
& \cmark\ \textbf{Addressed}
& \cmark\ \textbf{Addressed}
& \cmark\ \textbf{Explicit} \\

\bottomrule
\end{tabularx}

\caption{Comparison of existing survey traditions. The table contrasts prior surveys and this work by whether they explicitly address scientific epistemic context, human-role framing, and failure-mode analysis. These dimensions highlight the analytical gap addressed by CSCW$\times$Kuhn.}
\label{tab:survey_gap_comparison_full}
\end{table*}

\begin{table*}[t]
\centering
\scriptsize
\setlength{\tabcolsep}{5pt}
\renewcommand{\arraystretch}{1.12}

\begin{tabularx}{\textwidth}{
>{\raggedright\arraybackslash}p{0.18\textwidth}
>{\raggedright\arraybackslash}p{0.23\textwidth}
>{\raggedright\arraybackslash}X
}
\toprule
\textbf{Conceptual Block} &
\textbf{Term} &
\textbf{Operational Definition} \\
\midrule

\rowcolor{gray!12}
\multicolumn{3}{l}{\textbf{Human--Agent Scientific Collaboration}} \\
\midrule

Collaboration scope
&
\textbf{Human--Agent Collaboration}
&
A collaborative arrangement in which humans and agentic systems jointly contribute to a scientific goal. We exclude simple retrieval tools and static classifiers, and include agents that participate in planning, judgment, execution, or revision within scientific workflows.
\\

&
\textbf{Agentic system}
&
A system that integrates one or more AI agents capable of planning, reasoning, tool use, memory, workflow execution, or autonomous task decomposition.
\\

Coordination structure
&
\textbf{Human--agent coordination}
&
The organization of interdependent work between humans and agents, including task allocation, state tracking, feedback, and intervention.
\\

&
\textbf{Agent--agent coordination}
&
The organization of interdependent work among agents in multi-agent systems, including role assignment, communication, delegation, and aggregation.
\\

\midrule
\rowcolor{gray!12}
\multicolumn{3}{l}{\textbf{Coordination Dimensions}} \\
\midrule

Coordination demand
&
\textbf{Coordination demand}
&
The type and intensity of coordination required by a scientific workflow under a given epistemic context.
\\

&
\textit{Articulation work} \((\alpha)\)
&
The work required to organize, assign, sequence, integrate, and revise interdependent tasks so that the workflow remains coherent.
\\

&
\textit{Common ground} \((\gamma)\)
&
The shared understanding of goals, assumptions, constraints, evidence, methods, task state, and responsibilities among collaborators.
\\

&
\textit{Awareness} \((\omega)\)
&
The visibility of actions, intermediate outputs, decisions, errors, provenance, and consequences during collaboration.
\\

\midrule
\rowcolor{gray!12}
\multicolumn{3}{l}{\textbf{Scientific Context}} \\
\midrule

Scientific activity
&
\textbf{Scientific innovation}
&
The advancement of scientific knowledge through research, discovery, validation, and refinement of methods or explanations.
\\

&
\textit{Scientific research}
&
Systematic inquiry that formulates questions, applies methods, analyzes evidence, and produces scientific claims, models, tools, or artifacts.
\\

&
\textit{Scientific discovery}
&
The production of new, externally verifiable scientific knowledge or explanations.
\\

Epistemic condition
&
\textbf{Epistemic context}
&
The scientific conditions that shape collaboration, including uncertainty, assumptions, goals, evidential standards, and validation criteria.
\\

&
\textbf{Anomaly}
&
A finding, observation, or claim that does not fit prevailing assumptions and raises uncertainty about whether existing explanations remain sufficient.
\\

\midrule
\rowcolor{gray!12}
\multicolumn{3}{l}{\textbf{Epistemic Modes}} \\
\midrule

&
\textbf{Epistemic mode}
&
A form of scientific work characterized by distinct goals, uncertainty levels, evidential standards, and coordination demands.
\\

&
\textit{Puzzle-solving in normal science}
&
Routine scientific work conducted within accepted paradigms, methods, assumptions, and evaluation criteria.
\\

&
\textit{Anomaly evaluation}
&
The assessment of unexpected findings, uncertain claims, or contested contributions to determine their credibility, significance, and relation to existing assumptions.
\\

&
\textit{Paradigm-shift exploration}
&
Exploratory work that searches for alternative concepts, frameworks, or research directions when existing assumptions may be insufficient.
\\

\midrule
\rowcolor{gray!12}
\multicolumn{3}{l}{\textbf{Coordination Failure Modes}} \\
\midrule

&
\textbf{Failure mode}
&
A recurring breakdown associated with misalignment between coordination structure and epistemic demands in human--agent scientific collaboration.
\\

&
\textit{Procedural drift}
&
Deviation from the intended workflow, protocol, constraint set, or task state.
\\

&
\textit{Grounding substitution}
&
Replacement of evidential support or domain validity with plausibility, fluency, or apparent novelty.
\\

&
\textit{Evaluative flattening}
&
Reduction of contested scientific judgment into simplified scores, rankings, criteria, or majority opinions.
\\

&
\textit{Proxy optimization}
&
Optimization of benchmark metrics or proxy objectives while losing alignment with the underlying scientific goal.
\\

&
\textit{Epistemic black-boxing}
&
Limited inspectability into provenance, assumptions, rationale, intermediate states, tool use, or responsibility.
\\

\bottomrule
\end{tabularx}

\caption{Operational definitions of key concepts used in this survey. Terms are grouped by conceptual function to clarify how CSCW$\times$Kuhn connects human--agent collaboration, coordination demand, epistemic context, and coordination failure.}
\label{tab:operational-definitions}
\end{table*}

\newcolumntype{L}[1]{>{\raggedright\arraybackslash}p{#1}}
\newcolumntype{M}{>{\raggedright\arraybackslash}X}
\newcolumntype{Z}{>{\centering\arraybackslash}p{0.018\linewidth}}

\begin{table*}[p]
\centering
\tiny
\setlength{\tabcolsep}{2.5pt}
\renewcommand{\arraystretch}{1.02}

\begin{adjustbox}{width=\linewidth}
\begin{tabularx}{\linewidth}{
L{0.16\linewidth}
L{0.05\linewidth}
Z Z
L{0.35\linewidth}
M
Z Z Z Z
}
\toprule
\textbf{Research Work} &
\textbf{Science Domain} &
\multicolumn{2}{c}{\textbf{Scope}} &
\textbf{Target Task} &
\textbf{Metric / Approach} &
\multicolumn{4}{c}{\textbf{Coverage Gap}} \\
\cmidrule(lr){3-4}\cmidrule(lr){7-10}
& & \textbf{Seg.} & \textbf{Task} & & & \textbf{C} & \textbf{O} & \textbf{J} & \textbf{V} \\
\midrule

\multicolumn{10}{c}{\textbf{Execution and Automated Workflow}} \\
\midrule

MLE-Bench [B] \citep{chan2025mlebench}
& AI / ML
& \cmark
&
& Train models and submit prediction files for Kaggle competitions
& Any-medal rate; medal rates; above-median rate; valid submission rate; pass@k; raw competition score
& & & \xmark & \\

DiscoveryBench [B] \citep{Majumder2025DiscoveryBench}
& General
& \cmark
&
& Search and verify hypotheses from datasets and discovery goals
& Hypothesis Matching Score; context F1; variable F1; relationship accuracy
& \xmark & \xmark & & \\

ScienceAgentBench [B] \citep{Chen2025ScienceAgentBench}
& General
&
& \cmark
& Generate self-contained Python code to solve tasks from published papers
& Valid execution rate; success rate; CodeBERTScore; API cost; rubric-based human rating
& & & \xmark & \\

Auto-Bench [B] \citep{chen2025autobench}
& General
&
& \cmark
& Infer causal graphs through iterative oracle-guided interventions
& Success rate; average iterations; adjacency-matrix match; trajectory accuracy
& & \xmark & & \\

DSBench [B] \citep{jing2025dsbench}
& Data Science
&
& \cmark
& Answer data-analysis questions and build predictive models
& Task-level accuracy; competition-level accuracy; task success rate; performance gap
& & \xmark & \xmark & \\

AAAR-1.0 [B] \citep{Lou2025AAAR}
& General
&
& \cmark
& Validate equations, design experiments, detect paper weaknesses, identify human-review deficiencies
& F1; entailment precision/recall/F1; semantic similarity; ROUGE/BERTScore; TF-IDF
& \xmark & & & \\

RE-Bench [B] \citep{Wijk2025REBench}
& AI / ML
& \cmark
&
& Solve open-ended ML research-engineering environments
& Normalized score; score@k; time budget; human-expert comparison
& \xmark & & & \\

Gravity-Bench-v1 [B] \citep{koblischke2025gravitybench}
& Physics
&
& \cmark
& Infer gravitational dynamics from budgeted observations in simulated binary-star systems
& Task success rate; percent-error threshold; expert-reference comparison; observations used
& \xmark & & & \\

MLAlgo-Bench [B] \citep{wang2025mlalgobench}
& AI / ML
&
& \cmark
& Implement ML algorithms and workflows from instructions
& Pass rate; performance; time overhead; instruction-following score
& & & \xmark & \\

PaperBench [B] \citep{starace2025paperbench}
& AI / ML
& \cmark
&
& Replicate AI research papers
& Replication Score; rubric pass/fail scores; JudgeEval F1; human baseline comparison
& & \xmark & \xmark & \\

MLGym-Bench [B] \citep{nathani2025mlgym}
& AI / ML
& \cmark
&
& Improve baselines on open-ended AI research tasks
& AUP score; performance profiles; Best Attempt@4; Best Submission@4; raw task metrics
& & \xmark & \xmark & \\

SciReplicate-Bench [B] \citep{xiang2025scireplicatebench}
& NLP
&
& \cmark
& Generate code from algorithm descriptions in recent NLP papers
& Execution accuracy; CodeBLEU; reasoning graph accuracy; dependency/API recall
& & & \xmark & \\

\midrule
\multicolumn{10}{c}{\textbf{Scholarly Judgment and Evidential Support}} \\
\midrule

AgentReview [E] \citep{Jin2024AgentReviewEP}
& AI / ML
& \cmark
&
& Simulate peer-review process
& Decision agreement; Jaccard index; Cohen's \(\kappa\); rating mean/variance; review-text length; BERTScore/embedding similarity
& & & \xmark & \\

TruthHypo [B] / KnowHD [E] \citep{xiong2025truthhypo}
& Biomed.
&
& \cmark
& Generate and verify biomedical entity-relation hypotheses
& Link-level precision/recall/F1; relation accuracy; groundedness score; human/GPT-4o truthfulness preference
& & & \xmark & \xmark \\

SchNovel [B] / RAG-Novelty [E] \citep{lin2025schnovel}
& General
&
& \cmark
& Compare paired papers to identify the more novel work
& Pairwise accuracy; field-level accuracy; year-gap accuracy; RAG-Novelty accuracy
& & & \xmark & \\

ChemRAG-Bench [B] \citep{zhong2026chemragbench}
& Chemistry
&
& \cmark
& Evaluate RAG on chemistry question-answering, calculation, molecule, and reaction tasks
& Accuracy; exact match; validity; molecular fingerprint similarity; BLEU/ROUGE; error accuracy
& & & & \xmark \\

MLRC-Bench [B] \citep{zhang2025mlrcbench}
& AI / ML
& \cmark
&
& Propose and implement novel ML methods for research competitions
& Improvement to human solution; effectiveness; efficiency; code simplicity; pass@k
& & & \xmark & \\

Wu et al. (2025) [E] \citep{wu2025novelty}
& AI / ML
&
& \cmark
& Predict method novelty from novelty sentences in reviews and LLM-generated method summaries
& Accuracy; weighted F1
& & & \xmark & \\

CLAIMCHECK [B] \citep{ou2025claimcheck}
& AI / ML
&
& \cmark
& Analyze and verify review weaknesses in disputed paper claims
& Claim-association F1; weakness-label agreement; specificity/groundedness preference; weakness similarity
& & \xmark & \xmark & \\

ScholarQABench [B] / OpenScholar [E] \citep{asai2026openscholar}
& General
& \cmark
&
& Retrieve relevant literature and synthesize citation-backed answers to scientific queries
& Rubric accuracy; citation F1; correctness accuracy; ROUGE-L; LLM-judge quality scores; expert preference
& \xmark & & & \xmark \\

PaperWritingBench [B] \citep{song2026paperorchestra}
& AI / ML
& \cmark
&
& Generate LaTeX manuscripts, synthesize targeted related work, construct citations and visuals
& Citation F1/recall; literature-review/manuscript quality; simulated acceptance; human pairwise preference
& \xmark & & & \xmark \\

\midrule
\multicolumn{10}{c}{\textbf{Exploratory Ideation and Research-Direction Value}} \\
\midrule

MOOSE [B/E] \citep{yang2024opendomainhypotheses}
& Social Science
& \cmark
&
& Generate hypotheses from raw web corpus
& Validness score; novelty score; helpfulness score; expert ratings; feedback-iteration gains
& & & & \xmark \\

Si et al. (2025) [E] \citep{si2025llms}
& NLP
&
& \cmark
& Compare novelty in LLM-generated versus expert-written research ideas through blind expert review
& Novelty score; excitement score; feasibility score; expected-effectiveness score; overall score; reviewer agreement
& & & & \xmark \\

MOOSE-Chem [B/E] \citep{yang2025moosechem}
& Chemistry
& \cmark
&
& Rediscover chemistry hypotheses by inspiration retrieval, composition, and ranking
& Hit ratio; Matched Score; average rank ratio; Top MS; Average MS
& & \xmark & & \xmark \\

AI Idea Bench 2025 [B] \citep{qiu2025aiideabench}
& AI / ML
& \cmark
&
& Generate research ideas using target and inspiration papers
& Idea-to-topic score; idea-to-idea score; multiple-choice success; idea-competition score; novelty score; feasibility score; feasibility-per-step score
& & & \xmark & \xmark \\

IdeaBench [B] \citep{guo2025ideabench}
& General
& \cmark
&
& Generate research ideas from target-paper reference abstracts
& Insight Score; semantic similarity; idea-overlap rating; human-ranking alignment
& & & & \xmark \\

LiveIdeaBench [B] \citep{Ruan2026LiveIdeaBench}
& General
&
& \cmark
& Generate scientific ideas from single-keyword prompts
& Originality; feasibility; fluency; flexibility; clarity; overall score
& & & \xmark & \xmark \\

ResearchBench [B] \citep{liu2026researchbench}
& General
& \cmark
&
& Retrieve inspirations, compose hypotheses, and rank hypotheses
& Hit ratio; normalized hypothesis-composition score; pairwise ranking accuracy
& & \xmark & & \xmark \\

SDE [B/E] \citep{song2026sde}
& General
& \cmark
&
& Answer scenario-grounded questions; generate, test, and refine hypotheses in discovery loops
& Question-level accuracy; normalized project-level subobjective scores
& & & \xmark & \xmark \\

\bottomrule
\end{tabularx}
\end{adjustbox}

\vspace{2pt}
\caption{Benchmark and evaluation studies for human--agent scientific collaboration. 
[B] denotes benchmark and [E] denotes evaluation study. Seg. denotes workflow-segment evaluation, and Task denotes task-specific evaluation. 
Coverage gaps are marked for workflow coordination (C), human oversight (O), scientific judgment (J), and downstream validation (V).}
\label{tab:benchmark-eval-mapping}

\end{table*}

\end{document}